\documentclass[twocolumn,10pt]{article}

\usepackage[utf8]{inputenc}
\usepackage[T1]{fontenc}
\usepackage{lmodern}
\usepackage[margin=1.8cm,columnsep=0.6cm]{geometry}
\usepackage{graphicx}
\usepackage{amsmath,amssymb}
\usepackage{booktabs}
\usepackage{hyperref}
\usepackage{xcolor}
\usepackage{listings}
\usepackage{algorithm}
\usepackage{algorithmic}
\usepackage{caption}
\usepackage{subcaption}
\usepackage{siunitx}
\usepackage{tabularx}
\usepackage{url}
\usepackage{cite}
\usepackage{fancyhdr}
\usepackage{titlesec}
\usepackage{microtype}
\usepackage{needspace}
\usepackage{placeins}
\usepackage{float}
\usepackage{stfloats}

\hypersetup{
  colorlinks=true,
  linkcolor=blue!60!black,
  citecolor=green!50!black,
  urlcolor=blue!70!black,
}

\title{%
  \textbf{Rank-Aware CPU Reservation and In-Place Adjustment\\
  for MPI CFD on Kubernetes}%
}

\author{%
  Tianfang Xie\thanks{Corresponding author. E-mail: tianfangxie@purdue.edu}\\
  \textit{School of Aeronautics and Astronautics, Purdue University,}\\
  \textit{West Lafayette, IN 47907, USA}%
}

\date{}

\begin{document}
\maketitle

\begin{abstract}
Non-uniform meshes in computational fluid dynamics (CFD) can leave some Message Passing Interface (MPI) ranks waiting while others determine execution time.  On Kubernetes, reserving a full CPU for every rank can then restrict the admission of further simulations.  We investigate rank-aware CPU reservation and in-place request adjustment using OpenFOAM on three Kubernetes testbeds.  The framework lowers reservations without hard CPU caps and uses online MPI profiles to adjust each rank's CPU request while the solver runs.  We reduced the per-job CPU reservation from 16.0 to 5.9\,vCPU, with single-job wall-clock time increasing by at most 6\%.  With these reservations on twelve nodes, six co-located simulations reached 4.09 times the throughput of sequential execution.  Matched controls showed no consistent benefit from proportional request shaping.  Beyond the nominal CPU capacity, throughput collapsed with default busy-waiting; yielding waits softened the collapse.  In one six-simulation run, the online controller performed 120 in-place resizes of running solvers with zero pod restarts; the pipeline reached 2.5 times the throughput of sequential execution.  In a separate comparison on 16 physical cores, reallocating ranks from the far field to the near-wall zone reduced fixed-iteration wall-clock time by factors of 2.0 to 4.2 relative to the original concentric allocation, depending on mesh and iteration horizon.  The reallocation also outperformed balanced Scotch partitions.  The implementation requires no changes to OpenFOAM, the MPI library, or the Kubernetes scheduler.
\end{abstract}

\textbf{Keywords:} Kubernetes; Message Passing Interface (MPI); CPU reservation; CPU multiplexing; in-place pod resizing; domain decomposition; OpenFOAM; computational fluid dynamics

\section{Introduction}
\label{sec:introduction}

Computational Fluid Dynamics (CFD) simulations of turbulent flow solve the governing equations on meshes of millions of cells, often over thousands of iterations with implicit pressure--velocity coupling \cite{jasak2007openfoam,weller1998openfoam}.  The mesh is partitioned into subdomains, each advanced by a Message Passing Interface (MPI) rank that exchanges boundary data with its neighbors at every outer iteration \cite{mpi31standard,gropp1999using}.  Beyond the cost of one simulation, parameter sweeps, design-space exploration, and uncertainty quantification require tens to hundreds of independent cases.  We focus on cases of moderate size (4--32 ranks, $10^5$--$10^6$ cells) run on shared, cloud-managed clusters.

On these clusters, simulations and other services compete for a fixed pool of virtual CPUs (vCPUs) \cite{netto2018hpccloud,sochat2025usability}.  Under a reservation budget, capacity is accounted to its owner whether or not it is used.  The computational work can nevertheless vary across ranks.  A C-mesh around an airfoil is refined near the wall and coarser in the far field.  In the concentric partitions studied here, the near-wall subdomains hold an order of magnitude more cells than the far-field ones.  Collective synchronization makes lighter ranks wait for heavier ranks, while reserving a full CPU for each rank can limit the admission of further simulations.  Reducing this reservation therefore requires assessing both its single-job cost and the throughput of co-located simulations.

Kubernetes (K8s) \cite{kubernetes2024docs} provides CPU requests at milli-core granularity, allowing different reservations for individual ranks when each runs in its own pod.  Requests determine pod placement and set the CPU weights used under contention by the Linux fair scheduler \cite{turner2010cfs,lozi2016linux} through cgroups~v2 \cite{heo2023cgroups}.  Without hard CPU limits, they let a rank use available capacity during computation.  In-Place Pod Vertical Scaling (KEP-1287) \cite{kep1287,k8s135release}, stable since Kubernetes~v1.35, also allows requests to be changed while the pod runs, without restarting the container or losing the in-memory solver state.

The scope for CPU multiplexing depends on how co-located ranks compete for resources.  With the default busy-polling used here, an MPI rank remains runnable while waiting.  Studies of MPI oversubscription show that time-sharing can improve throughput when workloads have complementary demands, but can degrade performance when progress depends on frequent synchronization \cite{iancu2010oversubscription,utrera2004malleability}.  Prior MPI-on-Kubernetes evaluations have used uniform CPU requests across ranks \cite{beltre2019enabling,liu2022scanflow,sochat2025usability}.  Work on MPI elasticity changes rank counts through checkpoint-restart \cite{medeiros2024kub}, while ARC-V applies in-place resizing to memory provisioning \cite{medeiros2025arcv}.  These studies leave open how to set and adjust per-rank CPU requests during tightly coupled CFD runs, and what reduced reservations deliver when simulations share a cluster.

This paper develops a framework for rank-aware CPU reservation and in-place request adjustment for MPI CFD simulations on Kubernetes.  It uses per-rank load information to set CPU requests and updates them while the solver runs.  We evaluate the single-job cost of reservation changes on a four-node cluster, co-location throughput on a twelve-node cluster, and online adjustment driven by PMPI profiles of the production solver.  The main contributions are:

\begin{enumerate}
  \item \textbf{Rank-aware CPU reservation for MPI CFD.}  The framework maps the cell counts or the PMPI duty cycles of the ranks to CPU requests at milli-core granularity, so that sub-core and heterogeneous reservations are expressed through standard Kubernetes resource fields without changing the partition or the rank count.  Requests are applied without hard CPU limits, so that a rank can use available CPU capacity during computation (Section~\ref{sec:methodology}). Same-session controls separate what the reservation changes from the effects of the decomposition and of hard limits (Section~\ref{sec:res_single}).

  \item \textbf{CPU multiplexing under reduced reservations.} Reducing a job's reservation from 16.0 to 5.9\,vCPU increased single-job time by at most 6\%.  With this reduced reservation, six simulations on twelve nodes achieved 4.09$\times$ the throughput of the corresponding sequential reference.  Equal and uniform controls showed no consistent advantage of proportional requests, and the experiments identify the concurrency boundary beyond which busy-waiting degrades throughput, which motivates the check on the number of worker pods during admission (Section~\ref{sec:res_duty}).

  \item \textbf{In-place CPU request adjustment during solver execution.}  An online controller profiles the production solver and adjusts its per-rank CPU requests without a separate profiling run.  It preserves the MPI processes and their in-memory solver state, and admits further simulations subject to checks on reserved capacity and on the number of worker pods.  In one six-simulation pipeline, it applied 120 resizes to running solvers with zero pod restarts and reached 2.5$\times$ the throughput of the sequential reference.  The implementation requires no changes to OpenFOAM, the MPI library, or the Kubernetes scheduler (Section~\ref{sec:res_dynamic}).
\end{enumerate}

Domain decomposition determines how mesh cells are distributed among MPI ranks \cite{karypis1998metis,pellegrini1996scotch}, thereby defining the work that each rank's reservation must accommodate. A complementary 16-core experiment examines how rank allocation across mesh zones affects single-job time. Giving the near-wall zone more ranks and the far field fewer can reduce the heaviest ranks' work, while partition geometry can also affect pressure-solver convergence. This comparison uses different layouts from the reservation experiments and is evaluated independently.

Section~\ref{sec:background} reviews domain decomposition and MPI synchronization, CPU management on Kubernetes, related work, and the gap this paper addresses. Section~\ref{sec:methodology} defines the rank-level resource model, how load information becomes a CPU reservation, CPU multiplexing and admission control, and in-place request adjustment. Section~\ref{sec:setup} describes the three testbeds, cases, and comparisons, including the rank-allocation comparison at a fixed core budget. Section~\ref{sec:results} reports the results and discusses their implications and limitations, and Section~\ref{sec:conclusions} concludes.

\section{Background and Related Work}
\label{sec:background}

\subsection{Domain Decomposition and MPI Synchronization}
\label{sec:bg_openfoam}

OpenFOAM \cite{jasak2007openfoam,weller1998openfoam,openfoam2024guide} is an open-source finite-volume framework widely used for incompressible and compressible flow, combustion, multiphase, and solid-mechanics problems. Parallel execution follows a spatial decomposition approach: the utility \texttt{decomposePar} partitions the computational mesh into $n_r$ subdomains, each written to a separate directory (\texttt{processor0}, \texttt{processor1}, \ldots).  At runtime, $n_r$ MPI ranks are launched, each loading its local subdomain and communicating face-interpolated boundary values through MPI point-to-point and collective operations.

The partitioning algorithm is specified in the \texttt{decomposeParDict} dictionary.  Among the available methods, \texttt{scotch} \cite{pellegrini1996scotch} and \texttt{metis} \cite{karypis1998metis} employ graph-based multilevel $k$-way partitioning to minimize inter-processor communication while balancing cell counts.  Both methods accept an optional \texttt{processorWeights} list, a vector of $n_r$ floating-point values that biases the partitioner to assign proportionally more cells to subdomains with higher weights.  For example, a weight vector $(1, 1, 5, 15)$ normalized to unit sum instructs Scotch to place approximately $1/22$, $1/22$, $5/22$ and $15/22$ of the total cells in the four respective subdomains.  OpenFOAM also accepts a \texttt{manual} method in which every cell is explicitly assigned to a processor, allowing partitions to follow prescribed geometric boundaries.

\label{sec:bg_mpi}
The Message Passing Interface (MPI) \cite{mpi31standard} is a standardized API for distributed-memory parallel programming.  OpenFOAM uses MPI for all inter-process communication, relying on both point-to-point exchanges (\texttt{MPI\_Send}/\texttt{MPI\_Recv}) across processor boundaries and collective operations (\texttt{MPI\_Allreduce}, \texttt{MPI\_Allgather}) for global residual norms and convergence checks.  Iterative CFD solvers follow a bulk-synchronous pattern: a compute sweep over the local subdomain is followed by collectives at which every rank must arrive before any can proceed, so the job's iteration time is set by its slowest rank. Hoefler et al.\ \cite{hoefler2010characterizing} showed that even microsecond-scale operating-system noise on one rank amplifies across a communicator, and Petrini et al.\ \cite{petrini2003missing} traced a factor-of-two loss on ASCI~Q to exactly this straggler mechanism.  Any scheduler that delays one rank therefore delays the job.

The waiting mode determines whether synchronization leaves a CPU idle. Open~MPI \cite{gabriel2004openmpi}, in the default configuration used here, waits at barriers and in collectives by \emph{busy-polling} its progress engine (\texttt{mpi\_yield\_when\_idle=0}): a rank that has nothing to do consumes a full core with spin cycles until its peers arrive.  From the operating system's point of view, a waiting rank is indistinguishable from a computing one; it is runnable and uses CPU.

Container isolation also affects the communication path.  In the deployment used here, ranks of one job run in separate containers with separate IPC namespaces.  They cannot use the shared-memory transport (\texttt{vader}) between them even on the same node; intra-node traffic then travels over TCP through the loopback or pod network.

MPI implementations permit \emph{oversubscription}, launching more ranks than cores, in which case the operating-system scheduler time-slices ranks onto cores.  The behavior of oversubscribed MPI is well studied \cite{iancu2010oversubscription,utrera2004malleability}: throughput can improve when the co-scheduled processes have complementary demands, and it degrades sharply for synchronization-bound codes because time-slicing a rank away at the wrong moment stalls every other rank at the next barrier.  Left to a general-purpose scheduler with equal priorities, oversubscribed ranks share cores regardless of load, and the busy-waiting of light ranks competes on equal terms with the computation of heavy ranks.

These mechanisms connect the mesh partition to CPU provisioning.  For the finite-volume RANS solvers used here, the turbulence model is solved in every cell, and the per-cell cost is nearly uniform.  Cell count is therefore a first-order proxy for per-rank work (Section~\ref{sec:cell_proxy}).  Balancing cell counts can balance a compute sweep, but the partition shape also affects the number of pressure-solver iterations per step (Section~\ref{sec:res_blocks}).  The concentric, distance-based partitions used here follow the airfoil geometry and have unequal cell counts.  Their lighter ranks reach synchronization points earlier and spend more time waiting.

For a given partition, rank-level load information can guide CPU reservation without changing cell distribution or the number of ranks executing the job. OpenFOAM's \texttt{processorWeights} specifies target cell shares; the separate fixed-core comparison in Section~\ref{sec:res_blocks} evaluates changes to rank allocation across zones. The reservation study in Section~\ref{sec:res_duty} examines the single-job cost and co-location throughput of reduced requests, including the role of MPI waiting behavior.

\subsection{CPU Resource Management in Kubernetes}
\label{sec:bg_k8s}

Kubernetes \cite{kubernetes2024docs} is a container orchestration platform that manages workloads across a cluster of nodes.  Each container in a pod declares \emph{resource requests} and \emph{resource limits} for CPU and memory.  These two quantities serve different purposes:

\begin{description}
  \item[Requests] inform the scheduler's bin-packing algorithm.  A pod is placed on a node only if the node has sufficient unrequested capacity. Requests also set the \emph{cpu.weight} parameter in the cgroup, controlling the proportional share of CPU time when the node is contended.
  \item[Limits] set a hard ceiling enforced by the CFS bandwidth controller. The kernel assigns each cgroup a \emph{quota} $q$ (in microseconds) within a \emph{period} $P$ (default \SI{100}{\milli\second}).  A container with a CPU limit of $L$ cores receives a quota
        \begin{equation}
          q = L \times P,
          \label{eq:cfs_quota}
        \end{equation}
        so that a limit of $0.25$ CPU yields $q = \SI{25}{\milli\second}$ per \SI{100}{\milli\second} period.  When the container exhausts its quota, the kernel \emph{throttles} it; it deschedules all its threads until the next period begins.
\end{description}

Kubernetes exposes CPU in units of \emph{milli-cores}: \texttt{250m} represents $0.25$ of a core.  This granularity lets a reservation be expressed below one core, which whole-core binding cannot; it does not divide a single thread's speed (Section~\ref{sec:meth_request_semantics}).  Under cgroups~v2 \cite{heo2023cgroups}, the relevant control files are \texttt{cpu.max} (quota and period) for limits and \texttt{cpu.weight} for proportional sharing.  Kubernetes cannot express a limit below the request; when present, a limit is at least the request.  In the kernel used here (6.8, Table~\ref{tab:environment}), the fair scheduling class that consumes these weights selects tasks with the EEVDF algorithm, which replaced the CFS core in Linux~6.6 \cite{linux_eevdf}; \texttt{cpu.weight} keeps its proportional-share meaning, and the quota mechanism keeps its name, CFS bandwidth control.

Quality-of-Service (QoS) classes summarize the request and limit settings.  A pod whose containers all set CPU and memory limits equal to their requests is in the \emph{Guaranteed} class; its CPU limits apply even on an idle node.  A pod whose requests do not meet this condition is \emph{Burstable}.  It may have a CPU limit above its request or no CPU limit, and can use spare capacity up to any such limit.  For the shared-CPU configuration used here, requests without limits reserve capacity for scheduling while allowing a pod to use otherwise idle CPU time.  Weights affect the service received only when runnable threads contend for a logical CPU.  A rank that busy-waits at a barrier (Section~\ref{sec:bg_mpi}) remains runnable and competes for that service.  Section~\ref{sec:results} compares limits equal to requests with requests alone.

\label{sec:bg_inplace}
Historically, modifying a pod's resource requests or limits required deleting and recreating the pod, an unacceptable disruption for a running MPI job that holds the in-memory solver state.  Kubernetes Enhancement Proposal KEP-1287 \cite{kep1287} introduced \emph{In-Place Pod Vertical Scaling}, enabling the kubelet to resize a container's resources at runtime by updating the corresponding cgroup parameters without restarting the container. The feature entered alpha in Kubernetes~v1.27 (April 2023) behind the \texttt{InPlacePodVerticalScaling} feature gate, progressed to beta in v1.33 (April 2025) \cite{k8sinplacescaling}, and reached General Availability in v1.35 (December 2025) \cite{k8s135release}.  In GA, the feature gate is locked to \texttt{true} and the feature is enabled by default.  This makes CPU resizing available to running MPI jobs; the Kub system \cite{medeiros2024kub} deferred it while the feature was alpha (Section~\ref{sec:bg_related}). Each container declares a \texttt{resizePolicy} per resource; \texttt{NotRequired} means that the kubelet applies a change to that resource in place, which is the expected policy for CPU, since adjusting a cgroup weight or quota is a lightweight kernel operation, whereas memory may require \texttt{RestartContainer}. To resize a running pod, a client patches the \texttt{spec.containers[].resources} field through the Kubernetes resize API.  The kubelet detects the delta between desired and actual allocations, updates the cgroup parameters, and reports the new status in \texttt{status.containerStatuses[].resources}.  The resize is asynchronous; in the runs of Section~\ref{sec:res_dynamic} the new weight took effect within about 5\,s.

\subsection{Related Work}
\label{sec:bg_related}

The related work is organized by the layer at which resources are controlled: application-level load balancing and elasticity, co-scheduling and resource contention on shared hardware, and MPI resource management on Kubernetes.

\subsubsection{Application-Level Load Balancing and Elasticity}
Application-level methods redistribute work among MPI ranks.  ParMETIS \cite{schloegel2002parmetis}, Zoltan \cite{devine2002zoltan} and hypergraph repartitioners \cite{catalyurek2009repartitioning} migrate cells between ranks to balance their costs.  In OpenFOAM, Rettenmaier et al.\ \cite{rettenmaier2019openfoam} combined adaptive mesh refinement with dynamic load balancing.  Malleable and elastic MPI frameworks \cite{prabhakaran2015malleable,iserte2018dmr,compres2016elastic} change the rank count in cooperation with the batch system.  Houzeaux et al.\ \cite{houzeaux2022dynamic} applied this approach to the Alya CFD solver on Slurm, expanding or contracting the core allocation to hold a target parallel efficiency.  These methods require application support for data redistribution or rank-count changes, with associated checkpoint or migration overhead.

CPU reservation acts at a different level: it can be changed for each running rank while preserving the partition and rank count, and can therefore be combined with application-level repartitioning. The separate comparison in Section~\ref{sec:res_blocks} evaluates static rank allocation across zones alongside far-field core sharing at a fixed core budget. Weighted partitioning itself is already supported by Scotch, METIS, and OpenFOAM's \texttt{processorWeights}.

\subsubsection{Co-scheduling and Resource Contention}
Ousterhout \cite{ousterhout1982scheduling} established that communicating processes benefit from simultaneous scheduling to avoid cascading wait times.  Feitelson and Rudolph \cite{feitelson1992gang} formalized this as gang scheduling; Feitelson et al.\ \cite{feitelson1997theory} classified jobs as rigid, moldable or malleable, and Schwiegelshohn and Yahyapour \cite{schwiegelshohn2000fairness} analyzed fairness in parallel job scheduling.  Conventional batch allocations in systems such as Slurm \cite{yoo2003slurm} share a cluster between jobs in space rather than time: backfilling \cite{lifka1995anl,mualem2001backfill} lets small jobs run ahead in holes left by large ones, and node sharing allocates disjoint core sets of one node to different jobs.  Slurm can oversubscribe cores when configured to (\texttt{OverSubscribe}) and can shrink a running job's node allocation.  These mechanisms control job allocations rather than the per-rank CPU request interface studied here.

Iancu et al.\ \cite{iancu2010oversubscription} studied oversubscription of multicore processors directly and found that moderate oversubscription can improve system throughput when applications have complementary demands, but that synchronization-heavy codes degrade sharply.  This is consistent with the busy-wait collapse measured beyond the capacity boundary (Section~\ref{sec:res_capacity}). Utrera et al.\ \cite{utrera2004malleability} made MPI jobs malleable through folding, time-sharing several ranks of the same job on one core.  Section~\ref{sec:res_blocks} evaluates this folding for far-field CFD ranks using wall-clock time per pressure iteration. Section~\ref{sec:res_concurrent} instead co-locates ranks from different jobs, which rarely time-share inside the measured capacity boundary.  Zhuravlev et al.\
\cite{zhuravlev2010contention} classified schedulers that mitigate shared-resource contention on multicores.  Rabenseifner et al.\ \cite{rabenseifner2009hybrid} examined hybrid MPI+OpenMP to reduce per-node rank counts.  These approaches address placement and execution on shared cores; the Kubernetes request/limit interface adds a separate reservation decision.

Priedhorsky and Randles \cite{priedhorsky2017charliecloud} observed that HPC container runtimes deliberately avoid cgroup-based CPU isolation because it can harm MPI performance, a concern examined through throttle measurements in Section~\ref{sec:res_throttle}. Younge et al.\ \cite{younge2017containers} and Rudyy et al.\ \cite{rudyy2019containers} confirmed that container overhead is negligible for well-configured HPC workloads.

Co-location studies also examine shared-resource contention.  Mars et al.\ \cite{mars2011bubbleup} introduced Bubble-Up to predict interference between co-located workloads, finding memory bandwidth to be the dominant source of degradation for memory-intensive codes; Zhang et al.\ \cite{zhang2013cpi2} deployed contention detection based on cycles per instruction (CPI) at Google scale; Hager et al.\ \cite{hager2016exploring} showed with the execution-cache-memory (ECM) model that stencil-based PDE solvers are memory-bandwidth-bound; Lo et al.\ \cite{lo2015heracles} sustained 90\% utilization with hardware cache partitioning; and Delimitrou and Kozyrakis \cite{delimitrou2014quasar} automated co-location decisions from interference signatures.  These studies identify memory bandwidth as another potential limit on co-location density, discussed in Section~\ref{sec:discussion}.

\subsubsection{MPI Resource Management on Kubernetes}
Beltre et al.\ \cite{beltre2019enabling} provided the foundational evaluation of Kubernetes for MPI scientific workloads, benchmarking communication latency and HPCG throughput against bare metal on Chameleon Cloud.  They demonstrated near-bare-metal performance over RDMA but noted 13--22\% overhead for TCP/IP stacks.  All MPI ranks in their experiments receive identical CPU allocations; they do not explore per-rank resource differentiation.  Zhou et al.\ \cite{zhou2021container} proposed a hybrid Kubernetes--TORQUE architecture for HPC clusters, noting that Kubernetes scheduling is topology-unaware. Liu and Guitart \cite{liu2022scanflow} extended this direction with Scanflow-MPI, a two-layer scheduling framework that selects the number of containers per job, but still allocates uniform CPU resources across all containers.  Sochat et al.\ \cite{sochat2025usability} benchmarked 11 HPC applications on AWS, Azure, and GCP at up to 28\,672 CPUs; none of the applications is a CFD code, and they do not study per-rank weighting.

Verma et al.\ \cite{verma2015borg} and Burns et al.\ \cite{burns2016borg} described Google's Borg and the lessons it passed to Kubernetes, noting that safe overcommitment requires understanding per-application resource sensitivity; Rzadca et al.\ \cite{rzadca2020autopilot} demonstrated vertical autoscaling at Google scale for long-running services rather than batch MPI jobs.  Pe\~{n}a-Monferrer et al.\ \cite{penamonferrer2021cfd} ran concurrent CFD analysis and visualization pipelines on Kubernetes, focusing on post-processing rather than solver-level CPU allocation.

Medeiros et al.\ \cite{medeiros2024kub} introduced Kub, which adds \emph{horizontal} elasticity to MPI jobs on Kubernetes: at user-defined checkpoints, the running job is suspended, ranks are added or removed, data is reshuffled, and execution resumes.  They explicitly defer \emph{vertical} CPU scaling, adjusting CPU resources for already-running pods, as future work, citing the then-alpha status of the InPlacePodVerticalScaling feature. In a follow-on study, Medeiros et al.\ \cite{medeiros2025arcv} introduced ARC-V, which applies In-Place Pod Vertical Scaling to HPC and focuses exclusively on \emph{memory} provisioning across nine proxy applications; CPU vertical scaling is outside the scope of that work, and it considers no CFD workload.

\subsection{Research Gap and Scope}
\label{sec:bg_gap}

Table~\ref{tab:rw} distinguishes the capabilities covered by the prior work.  With one MPI rank per pod, the Kubernetes resource model can express fractional CPU reservations at milli-core granularity, different requests for individual ranks, weighted core sharing under contention, and in-place changes to each rank's request while the job runs.  These controls allow CPU reservations to be based on per-rank load information while preserving the solver state, partition, and rank count.

For tightly coupled CFD, lowering a job's reservation leaves more capacity available for admission, but co-location throughput depends on how ranks compete for CPU time. We develop a framework that uses rank-level load information to set requests and online profiles to update them while solvers run. The evaluation measures the single-job cost of reduced reservations and the throughput of co-located simulations, distinguishing the effects of reservation total, request shape, and MPI waiting behavior.

This rank-level control complements job-level admission.  Queuing systems such as Volcano and Kueue admit whole jobs according to their declared requests; the mechanism studied here determines those requests and updates them while ranks run.

\begin{table*}[t]
  \centering
  \small
  \caption{Positioning relative to key prior work}
  \label{tab:rw}
  \begin{tabular}{@{}lcccccc@{}}
    \toprule
    \textbf{Work} & \textbf{Year} &
    \textbf{K8s} & \textbf{CFD} &
    \textbf{Prop.\ CPU} & \textbf{Multi-job} & \textbf{In-place CPU} \\
    \midrule
    Iancu et al.\ \cite{iancu2010oversubscription} & 2010 & -- & -- & -- & \checkmark & -- \\
    Mars et al.\ \cite{mars2011bubbleup}            & 2011 & -- & -- & -- & \checkmark & -- \\
    Lo et al.\ \cite{lo2015heracles}                & 2015 & -- & -- & -- & \checkmark & -- \\
    Beltre et al.\ \cite{beltre2019enabling}        & 2019 & \checkmark & -- & -- & -- & -- \\
    Houzeaux et al.\ \cite{houzeaux2022dynamic}     & 2022 & -- & \checkmark & -- & -- & $\circ$ \\
    Liu \& Guitart \cite{liu2022scanflow}           & 2022 & \checkmark & -- & -- & -- & -- \\
    Medeiros et al.\ \cite{medeiros2024kub}         & 2023 & \checkmark & -- & -- & -- & -- \\
    Medeiros et al.\ \cite{medeiros2025arcv}        & 2025 & \checkmark & -- & -- & -- & mem.\ only \\
    \textbf{This work}                              & 2026 & \checkmark & \checkmark & \checkmark & \checkmark & \checkmark \\
    \bottomrule
  \end{tabular}
  \par\vspace{3pt}{\footnotesize\raggedright \checkmark = yes; -- = no; $\circ$ = present in a restricted form (for example per-job rather than per-rank CPU, or resizing of memory rather than CPU).\par}
\end{table*}

\section{Rank-Aware CPU Reservation and In-Place Adjustment}
\label{sec:methodology}

For a given MPI partition and rank count, the framework uses rank-level load information to set and update per-pod CPU requests. These requests determine the job's reservation and influence CPU sharing under contention, while the solver retains its partition, rank count, and in-memory state. The design connects three decisions: how to construct the request vector, when to admit another simulation, and how to revise requests during execution. Requests are applied without hard CPU limits. Admission checks both reservation headroom and the worker-pod count; the online controller uses a production-run PMPI profile to initialize requests and later makes bounded adjustments based on iteration rates.

\subsection{Framework Overview and Resource Model}
\label{sec:meth_model}

For a job with $n_r$ ranks, each rank runs in its own pod. The framework sets the CPU request $c_i$ of rank~$i$ in vCPU, which the Kubernetes scheduler accounts for during placement and the kubelet maps to a cgroup CPU weight (Section~\ref{sec:bg_k8s}). The request vector $(c_0, \ldots, c_{n_r-1})$ has two properties: its total $C_{\text{job}} = \sum_i c_i$, the job's reservation, and its distribution across ranks. The experiments vary these separately. CPU limits and MPI waiting behavior condition how the requests affect execution (Sections~\ref{sec:meth_limits} and~\ref{sec:theory}).

\begin{figure*}[!t]
  \centering
  \includegraphics[width=0.95\textwidth]{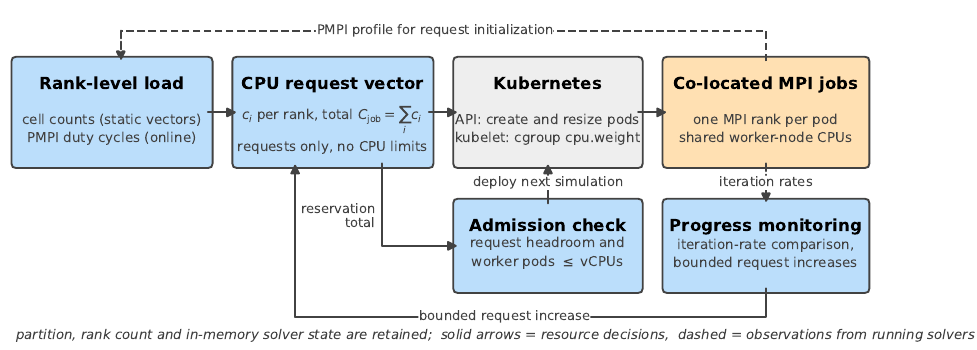}
  \caption{Framework for rank-level CPU reservation, co-location, and in-place adjustment. Cell counts define static request vectors; the online controller initializes requests from the production solver's PMPI profiles. Admission checks reservation headroom and the worker-pod count. Subsequent bounded request increases use iteration-rate comparisons. The partition, rank count, and in-memory solver state are retained. Solid arrows show resource decisions; dashed arrows show observations from running solvers.}
  \label{fig:control_loop}
\end{figure*}

Figure~\ref{fig:control_loop} connects the load information, resource decisions, and running solvers. Cell counts supply static request vectors; PMPI duty cycles supply the online controller's initial estimate. The controller applies requests through the Kubernetes resize interface, checks whether another simulation can be admitted, and uses solver iteration rates for subsequent adjustments. Section~\ref{sec:setup_metrics} defines the single-job timing, reservation, and throughput measurements. The independent rank-allocation comparison changes the partition supplied to the framework and is specified in Section~\ref{sec:meth_blocks}.

The mechanism requires no application-source or scheduler changes: profiling attaches through \texttt{LD\_PRELOAD}, and resource control uses standard pod fields interpreted by Kubernetes and the Linux fair scheduler. It applies to dynamically linked MPI applications that support preloading. Rank-level reservations can be combined with queue-level gang schedulers such as Volcano and Kueue, and with namespaces, \texttt{ResourceQuota}, and \texttt{PriorityClass} objects that bound and prioritize tenants.

\subsection{Load Estimation and CPU Reservation}
\label{sec:meth_proportional}

\begin{figure}[t]
  \centering
  \includegraphics[width=\columnwidth]{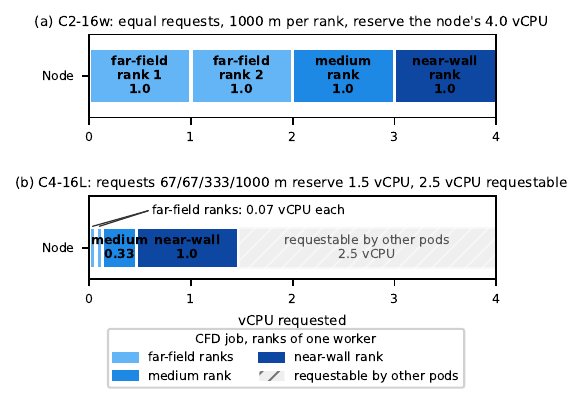}
  \caption{One c5.xlarge worker of Testbed~A under the measured 16-rank configurations; each worker hosts two far-field, one medium, and one near-wall rank. (a)~Equal requests of 1000\,m (C2-16w) reserve 4.0\,vCPU. (b)~Reduced proportional requests of 67/67/333/1000\,m (C4-16L) reserve 1.5\,vCPU and leave 2.5\,vCPU requestable by other pods (gray). Across four workers, the job reserves 5.9 of 16.0\,vCPU. Values are nominal and rounded; in panel~(a), rank~0 requests 60\,m less than nominal so that the launcher pod fits (Section~\ref{sec:setup_configs}).}
  \label{fig:packing}
\end{figure}

The first step is to translate per-rank load information into a CPU request vector. Cell counts provide a static proxy for a given partition; PMPI profiles provide a measurement-based alternative. Figure~\ref{fig:packing} illustrates the reservation reduction for the same rank placement under equal requests (C2-16w) and reduced proportional requests (C4-16L). The following subsections define the two load signals and how Kubernetes applies the resulting requests.

\subsubsection{Cell-Based Request Vectors}
\label{sec:cell_proxy}

For a given partition, the cell share of each rank supplies a first-order load estimate. OpenFOAM's \texttt{processorWeights} specifies target cell shares through a weight vector $\mathbf{w} = (w_0, w_1, \ldots, w_{n_r-1})$ for $n_r$ ranks:
\begin{equation}
  f_i = \frac{w_i}{\sum_{j=0}^{n_r-1} w_j}
  \label{eq:cell_fraction}
\end{equation}
Rank~$i$ receives approximately $f_i\,n_{\text{cell}}$ of the total $n_{\text{cell}}$ cells. A graph partitioner chooses the subdomain shapes; a manual assignment also prescribes their spatial membership. Section~\ref{sec:setup_configs} defines the weighted and geometric partitions used in the experiments.

Given a total request budget $C_{\text{job}}$ in vCPU, the proportional request of rank~$i$ is
\begin{equation}
  c_i = f_i \cdot C_{\text{job}} = \frac{w_i}{\sum_{j=0}^{n_r-1} w_j} \cdot C_{\text{job}}.
  \label{eq:cpu_alloc}
\end{equation}
The CPU-to-cell ratio $c_i / (f_i n_{\text{cell}}) = C_{\text{job}} / n_{\text{cell}}$ is then uniform. This construction distributes the reservation according to the cell-count proxy; Section~\ref{sec:meth_request_semantics} explains its execution semantics.

The following shape-only example keeps the total at the equal-request budget. For four ranks with weights $\mathbf{w} = (1, 1, 5, 15)$, the normalized fractions are $(1/22, 1/22, 5/22, 15/22)$. With $C_{\text{job}} = 4.0$ cores:
\begin{align}
  c_0 &= \tfrac{1}{22} \times 4.0 \approx 0.182 \;\text{CPU} \notag \\
  c_1 &= \tfrac{1}{22} \times 4.0 \approx 0.182 \;\text{CPU} \notag \\
  c_2 &= \tfrac{5}{22} \times 4.0 \approx 0.909 \;\text{CPU} \notag \\
  c_3 &= \tfrac{15}{22} \times 4.0 \approx 2.727 \;\text{CPU}
  \label{eq:example_alloc}
\end{align}
The requests are \texttt{182m}, \texttt{182m}, \texttt{909m}, and \texttt{2727m}; the helper script adjusts rounding so that $\sum c_i = C_{\text{job}}$. The reduced vectors in Fig.~\ref{fig:packing} also lower the total. The evaluation includes \emph{equal} requests of 1000\,m per rank and \emph{uniform} reduced requests with the same nominal total as the proportional vector (369\,m per rank for the 16-rank job). These controls separate the reservation total from its distribution.

Equation~\eqref{eq:cpu_alloc} assumes that cell count approximates per-rank computational cost. It is most appropriate when the solver performs similar work per cell, cell quality is comparable across subdomains, and rank-local features such as dynamic meshing or localized turbulence activation do not dominate the cost. The pitzDaily case uses a structured mesh with a uniformly applied $k$-$\varepsilon$ model; NACA~0012 uses a structured C-mesh with a uniformly applied $k$-$\omega$~SST model. The static vectors derive from cell counts: the four-rank $(1, 1, 5, 15)$ weights and the $1{:}5{:}15$ per-rank cell ratio of the 16-rank concentric partition (67/335/1005\,m). Their application and controls are specified in Section~\ref{sec:setup_configs}. The controller instead derives its initial vector from measured duty cycles (Section~\ref{sec:meth_dynamic}).

\subsubsection{PMPI Profiling and Duty-Cycle Estimation}
\label{sec:pmpi}
\label{sec:meth_idle}
\label{sec:duty-cycle}

\begin{sloppypar}
The PMPI profiling interface \cite{mpi31standard} intercepts MPI calls without modifying application source code. Our shared library, \texttt{libpmpi\_trace.so}, wraps seven communication and synchronization operations: \texttt{MPI\_Barrier}, \texttt{MPI\_Allreduce}, \texttt{MPI\_Alltoall}, \texttt{MPI\_Alltoallv}, \texttt{MPI\_Sendrecv}, \texttt{MPI\_Wait}, and \texttt{MPI\_Waitall}. It records rank indices and entry/exit timestamps through \texttt{clock\_gettime(CLOCK\_MONOTONIC)} at microsecond resolution, buffers them, and writes per-rank CSV traces and summaries at \texttt{MPI\_Finalize}. Optional window reports provide duty cycles while the solver runs. The library is loaded through \texttt{LD\_PRELOAD}, without recompiling OpenFOAM or the MPI runtime.
\end{sloppypar}

The \emph{CPU duty cycle} of rank~$i$ is the fraction of its run, from \texttt{MPI\_Init} to \texttt{MPI\_Finalize}, spent outside the intercepted calls:
\begin{equation}
  \bar{d}_i = 1 - \frac{t_{\text{MPI},i}}{t_{\text{wall},i}},
  \label{eq:duty-cycle}
\end{equation}
Here $t_{\text{wall},i}$ is elapsed wall-clock time and $t_{\text{MPI},i}$ is cumulative time inside those calls. The complement includes computation, file I/O, and unintercepted MPI calls, so $\bar{d}_i$ is an upper bound on useful computation. Correspondingly, $1 - \bar{d}_i$ is a lower bound on time inside MPI communication and synchronization, which includes message handling as well as waiting. The 200- and 50-iteration profiles include the startup transient.

Duty cycle differs from CPU occupancy: a rank spinning at a barrier can have low duty cycle and high occupancy. For applications that cannot be intercepted at the MPI layer, the framework can obtain occupancy from the change in the cgroup~v2 \texttt{cpu.stat} field \texttt{usage\_usec} divided by elapsed wall-clock time. This measures CPU time consumed rather than time outside MPI. Section~\ref{sec:res_duty} compares these signals; Section~\ref{sec:meth_dynamic} specifies how the controller converts PMPI duty cycles into requests.

\subsubsection{CPU Request Semantics and Enforcement}
\label{sec:meth_request_semantics}
\label{sec:meth_limits}

A request determines reservation and influences CPU service under contention. Each OpenFOAM rank here is single-threaded and occupies at most one logical CPU at a time: a request above 1000\,m cannot increase its simultaneous execution capacity, and a request below 1000\,m does not cap it when no limit is set. Requests act through three mechanisms:

\begin{enumerate}
  \item \textbf{Placement.} The scheduler places a pod on a node only if its unrequested capacity can accommodate the request $c_i$.
  \item \textbf{Weight under contention.} The kubelet maps $c_i$ to cgroup \texttt{cpu.weight} on a 1--10\,000 scale (1000\,m gives 39; 67\,m gives 3). These weights influence CPU time shares when runnable threads compete.
  \item \textbf{Load balancing.} Weighted load also influences how the Linux scheduler distributes threads across logical CPUs and simultaneous multithreading (SMT) siblings.
\end{enumerate}

The request vector therefore sets resource reservations and sharing weights, while the partition determines each rank's work. Reducing the total changes reservation headroom; redistributing it at the same total releases none. The single-job controls in Section~\ref{sec:res_single} distinguish these effects, and Section~\ref{sec:meth_multi} defines their role in co-location.

The framework uses \emph{requests only}, without hard CPU limits, so ranks can use available capacity during compute phases. The hard-limit controls instead set both \path{resources.requests.cpu} and \path{resources.limits.cpu} to $c_i$, applying the quota of Eq.~\eqref{eq:cfs_quota} with $L = c_i$. Computation and busy-waiting both consume the quota; exhaustion suspends the cgroup until replenishment. Section~\ref{sec:res_throttle} measures the resulting throttling for the sub-core limits tested here.

A limit at or above peak demand need not throttle a rank. Kubernetes permits a sub-core request with a full-core limit in a Burstable pod, whereas Guaranteed QoS requires matching requests and limits. Our experiments compare matching limits with no limits; limits above requests remain future work. Throttling is recorded through \texttt{nr\_throttled} and \texttt{throttled\_usec} in \texttt{cpu.stat}, available through cAdvisor or \texttt{/sys/fs/cgroup}.

\label{sec:pod-spec}
Listing~\ref{lst:pod-manifest} combines the request and resize policy for a sparse rank of simulation~A. The \texttt{sim} label identifies the job, the request sets its reservation and CPU weight, and \texttt{resizePolicy} allows CPU updates without restarting the container.
\begin{minipage}{\linewidth}
\begin{lstlisting}[
  caption={Pod specification for a sparse rank (67\,m CPU request), with in-place CPU resizing.},
  label={lst:pod-manifest}]
apiVersion: v1
kind: Pod
metadata:
  name: of-worker-a-0
  labels:
    app: openfoam
    role: worker
    sim: A
spec:
  containers:
  - name: openfoam
    image: openfoam-k8s:v10
    resources:
      requests:
        cpu: "67m"
      # No limits: Burstable QoS
    resizePolicy:
    - resourceName: cpu
      restartPolicy: NotRequired
\end{lstlisting}
\end{minipage}

The \texttt{67m} request is the sparse rank's share of a 1000\,m dense rank under the $1{:}5{:}15$ per-rank cell ratio (5.7\,K, 28\,K, and 85\,K cells per rank on the 499\,K mesh). A 16-rank job then reserves $8 \times 67 + 4 \times 333 + 4 \times 1000 \approx 5\,900$\,m $= 5.9$\,vCPU instead of 16.0\,vCPU. The deployed Testbed~B manifests round the medium and dense requests up to 335 and 1005\,m (Section~\ref{sec:setup_colocation}).

\subsection{CPU Multiplexing and Admission Control}
\label{sec:meth_multi}

Co-location means running several simulations on the same worker nodes; CPU multiplexing refers to sharing those nodes' CPUs among the simulations. Reduced requests leave more capacity available in reservation accounting. Efficient execution also depends on how many ranks compete for the CPUs and how they wait. We therefore distinguish reservation headroom from the worker-pod boundary used by the controller's admission check.

\subsubsection{Reservation Accounting and the Concurrency Boundary}
\label{sec:theory}

For a rank with steady-state duty cycle $\bar{d}_i$ and reservation $r_i$~millicores, define
\begin{equation}
  c^{\text{idle}}_i = r_i \cdot (1 - \bar{d}_i)
  \label{eq:reclaimable}
\end{equation}
as its \emph{reservation-weighted waiting}. The sum $\sum_i c^{\text{idle}}_i$ associates the job's reservation with time spent inside the intercepted MPI calls. It is an accounting quantity, not a measurement of CPU released: a busy-waiting rank still consumes CPU time. The duty-cycle and occupancy measurements in Section~\ref{sec:res_duty} illustrate the distinction, while the reservation/performance trade-off is evaluated in Sections~\ref{sec:res_scaling} and~\ref{sec:res_concurrent}.

With default busy-waiting, a rank remains runnable during both computation and MPI waits. If worker pods do not outnumber logical CPUs and placement is balanced, each worker can occupy a logical CPU; launcher and system activity can still compete for CPU time, and SMT siblings share physical-core resources. Once runnable threads outnumber logical CPUs, they must share CPU time, so waiting ranks' spin competes with computation. Yielding waits can reduce this competition, as tested in Section~\ref{sec:res_capacity}.

\begin{figure*}[!t]
  \centering
  \includegraphics[width=0.95\textwidth]{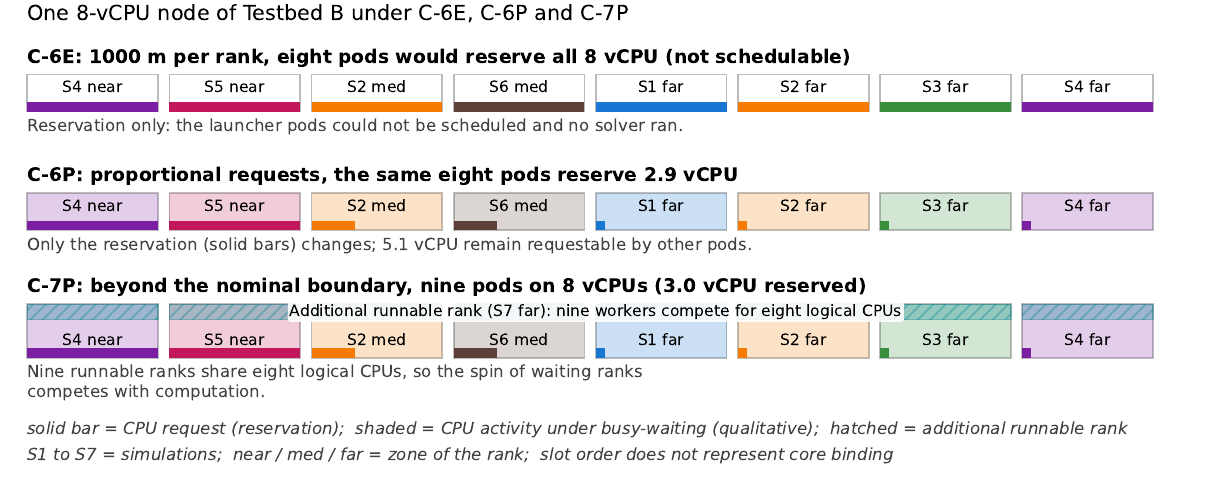}
  \caption{One 8-vCPU node of Testbed~B with the pods it hosted in the C-6P and C-7P runs, one slot per vCPU; the slot order does not represent core binding. At 1000\,m per worker (C-6E, top) the same eight pods would reserve all 8\,vCPU; the configuration could not be scheduled because the launcher requests did not fit, so only the reservation is drawn. Under the proportional vector (C-6P, middle) the eight pods reserve 2.948\,vCPU (solid bars); the shading indicates that busy-waiting ranks keep their vCPU active, as the occupancy measurements of Section~\ref{sec:res_waiting_profile} show, so further pods fit by request. Beyond the nominal boundary, the node hosted nine pods (C-7P, bottom, 3.015\,vCPU reserved): nine runnable ranks compete for eight logical CPUs (hatched band), so the spin of waiting ranks competes with computation.}
  \label{fig:mux_architecture}
\end{figure*}

The nominal concurrency boundary under balanced placement is
\begin{equation}
  N_{\max} = \left\lfloor \frac{M\,C_{\text{node}}}{n_r} \right\rfloor,
  \label{eq:capacity-boundary}
\end{equation}
where $M$ is the number of worker nodes, $C_{\text{node}}$ is vCPUs per node, and $n_r$ is ranks per simulation. This is the largest $N$ for which every worker pod can occupy its own vCPU. It depends on the deployment; Section~\ref{sec:res_capacity} tests the associated throughput boundary on the twelve-node cluster. Figure~\ref{fig:mux_architecture} separates reservation headroom from worker counts on one node.

\subsubsection{Throughput Model under CPU Contention}
\label{sec:model}

We model the makespan of $N$ co-located simulations, the completion time of the slowest one (Section~\ref{sec:setup_metrics}). Define normalized cluster load as $\rho_N = N n_r/(M C_{\text{node}})$. As this load approaches one, more simulations share the nodes' SMT siblings, memory systems, and network paths. A linear contention model describes the associated increase in makespan:
\begin{equation}
  T_N = T_1 \bigl(1 + \beta\,(\rho_N - \rho_1)\bigr),
  \label{eq:makespan-model}
\end{equation}
Here $T_1$ is the single-simulation makespan, $\rho_1 = n_r/(M C_{\text{node}})$ is the baseline load, and $\beta$ is a fitted contention coefficient that aggregates these sharing costs. The corresponding throughput gain is
\begin{equation}
  \Theta(N) = \frac{N\,T_1}{T_N}
            = \frac{N}{1 + \beta\,(\rho_N - \rho_1)}.
  \label{eq:throughput-model}
\end{equation}
The model applies within $\rho_N \le 1$; beyond that boundary, additional runnable ranks compete directly for logical CPU time. Section~\ref{sec:res_pareto} compares a fit over the measured range with predictions calibrated from a dual-simulation experiment alone. The coefficient is workload-dependent; its transfer across mesh resolutions is examined in the companion preprint \cite{xie2026multiplexing}. The controller uses the reservation and worker-count checks, rather than this fitted throughput model, for admission.

\subsubsection{MPI Deployment and Isolation}
\label{sec:k8s-implementation}
\label{sec:mpi-isolation}

Each simulation has $n_r$ worker pods and a dedicated launcher pod running \texttt{mpirun}; two 16-rank simulations therefore use 32~workers and two launchers. Job-specific \texttt{sim} labels and case directories on a shared Amazon Elastic File System (EFS) volume separate the solver outputs. Each launcher receives a ConfigMap hostfile containing only its job's pod IPs, one slot per pod. It launches processes through SSH with TCP transport and no explicit binding or mapping options. Yielding variants add \texttt{-{}-mca mpi\_yield\_when\_idle 1}. Workers use requests without limits; no affinity rules are imposed, and Kubernetes places pods across the shared nodes.

Separate pod IPC namespaces prevent Open~MPI's \texttt{vader} transport from connecting ranks on the same node. Jobs therefore use TCP over the pod network, selected explicitly with \texttt{-{}-mca btl tcp,self}. This separates shared-memory communication namespaces across jobs but leaves CPU, memory-system, and network contention. A shared IPC namespace through \texttt{hostIPC} is possible in Kubernetes and was not used in this study. Supplementary Section~S1 reports the configuration-specific pitzDaily communication-path check; Section~\ref{sec:res_waiting_profile} reports the Testbed~B host-network profile.

\subsection{In-Place CPU Request Adjustment}
\label{sec:meth_dynamic}
\label{sec:controller-design}

The online controller starts each simulation at equal requests, profiles the production solver, and resizes its pods without restarting the MPI ranks. It then checks admission of the next simulation and monitors progress to make bounded request increases. Kubernetes In-Place Pod Vertical Scaling supplies the resize interface; Listing~\ref{lst:pod-manifest} shows the CPU restart policy. The controller takes the case, rank count, and admission cap as inputs and executes the following phases.

\paragraph{Phase A, profile.} Obtain per-rank duty cycles $\bar{d}_i$ (Eq.~\eqref{eq:duty-cycle}). In the online design, the production solver starts with the PMPI wrapper attached and reports consecutive 60\,s windows. Phase~A reads the latest window after 240\,s. The pass-based comparison instead obtains duty cycles from a separate 50-iteration profiling run at equal requests before production.

\paragraph{Phase B, resize.} Set $r_i = r_{\text{base}} \cdot \bar{d}_i / \max_j \bar{d}_j$, with $r_{\text{base}} = 1000$\,m, and apply all $n_r$ changes through the in-place resize API (KEP-1287). The pods retain their case directories and network identities. The online production run continues through this update; in the pass-based design, production starts after profiling and initialization. Section~\ref{sec:res_dynamic} reports the resulting requests and resize latency.

\paragraph{Phase C, pack.} Check aggregate requests against allocatable cluster CPU and the worker-pod count against the vCPU count. Admit the next simulation only when request headroom passes the admission check and $n_r(N+1) \le M C_{\text{node}}$; then profile and resize it in turn. The two checks serve different purposes: checking aggregate requests alone would admit eight simulations under proportional requests, reserving only 47\,vCPU of 96. That concurrency exceeds the worker-count boundary in Eq.~\eqref{eq:capacity-boundary}. The evaluated controller uses a cap of six simulations (Section~\ref{sec:res_dynamic}).

\paragraph{Phase D, monitor.} Every minute, compare iteration rates among simulations that have been in production for at least two intervals. If the fastest-to-slowest rate ratio exceeds 1.15, increase the dense ranks' requests in the slowest simulation by 10\%, at most twice per simulation. Comparing rates avoids treating a later start as slower progress. This interval-based check differs from the run-level fairness metric in Section~\ref{sec:setup_metrics}. It uses solver logs; PMPI supplies the initial request estimate in Phases~A and~B.

\paragraph{Preset-phase mechanism check.}
\label{sec:meth_phase_scaling}
A separate \emph{pseudo-dynamic} protocol uses phase boundaries $t_1$ and $t_2$ fixed in advance, with requests $c_i^{\max}$, $c_i^{\mathrm{mid}}$, and $c_i^{\min}$:
\begin{equation}
  c_i(t) =
  \begin{cases}
    c_i^{\max}  & \text{if } t < t_1, \\
    c_i^{\mathrm{mid}} & \text{if } t_1 \le t < t_2, \\
    c_i^{\min}  & \text{if } t \ge t_2.
  \end{cases}
  \label{eq:dynamic_alloc}
\end{equation}
The update is applied to all worker pods at each transition. This auxiliary protocol checks whether running MPI ranks survive a resize without stall or data loss; it does not evaluate a measurement-driven policy. The C5 experiment exercises only the Phase~1$\to$2 transition (Sections~\ref{sec:setup_controller} and~\ref{sec:res_dynamic}).

\section{Experimental Setup}
\label{sec:setup}

\subsection{Hardware and Software}
\label{sec:setup_hw}

All experiments were conducted on clusters deployed on Amazon Web Services (AWS) EC2 and managed by k3s~v1.35, a lightweight, CNCF-certified Kubernetes distribution.  Three testbeds are used (Table~\ref{tab:environment}).  Testbed~A evaluates decomposition and CPU settings for a single simulation.  Testbed~B evaluates CPU multiplexing across co-located simulations and in-place CPU request adjustment.  Testbed~C provides a separate single-simulation comparison of rank allocation and core sharing.

\paragraph{Testbed~A (single-simulation experiments).}
This testbed consists of a control-plane node (\texttt{c5.large}) and two or four \texttt{c5.xlarge} worker nodes.  A \texttt{c5.xlarge} instance exposes 4~vCPUs backed by two physical cores with simultaneous multithreading (SMT), so that four single-threaded ranks on one node share two physical cores.  For 4-rank experiments, two workers each host two pods (one rank per vCPU); for 16-rank experiments, four workers each host four pods (one rank per vCPU, no oversubscription).  The c5 family is compute-optimized and non-burstable.

\paragraph{Testbed~B (CPU multiplexing and in-place adjustment).}
This testbed consists of a control-plane node (\texttt{t3.medium}, tainted \texttt{NoSchedule}) and twelve \texttt{c5.2xlarge} worker nodes, each with 8~vCPUs on four physical cores with SMT, \SI{16}{\gibi\byte} RAM and a single NUMA domain, giving 96 allocatable vCPUs.  A single 16-rank simulation places one or two ranks per node; $N$ concurrent simulations place $16N$ worker pods across the same twelve nodes, up to 128 pods at $N=8$.

\paragraph{Testbed~C (rank-allocation comparison).}
This testbed consists of a control-plane node (\texttt{t3.medium}) and one \texttt{c5.9xlarge} worker launched with SMT disabled, so that its 18 vCPUs are 18 physical cores and no rank shares a core with another through hyper-threading.  Sixteen cores form the core budget of every pinned configuration; the remaining two hold the system daemons and the MPI launcher pod.  Each rank is pinned to its core, or in the core-sharing configurations to a shared core, with \texttt{taskset} from a per-rank map, which is the placement the kubelet's static CPU manager gives Guaranteed pods with integer CPU requests.  One unpinned configuration leaves placement to the Linux fair scheduler; its ranks are then free to use all 18 cores (Section~\ref{sec:res_blocks}).

\paragraph{Burstable instances.}
The pitzDaily experiments (Section~\ref{sec:res_throttle} and Supplementary Section~S1) were run on a two-worker \texttt{t3.xlarge} cluster, a burstable instance type.  The resulting single-run timings are used only for qualitative comparisons.  Every NACA~0012 performance comparison runs on non-burstable c5 instances; the one NACA run on the t3 cluster, the preset-phase resize demonstration C5 (Section~\ref{sec:res_dynamic}), is not used for timing.

\begin{table*}[t]
  \centering
  \small
  \caption{Hardware and software environment of the three testbeds}
  \label{tab:environment}
  \begin{tabular}{@{}lp{3.6cm}p{3.6cm}p{3.6cm}@{}}
    \toprule
    \textbf{Component} & \textbf{Testbed A} & \textbf{Testbed B} & \textbf{Testbed C} \\
    \midrule
    Control plane      & 1$\times$ \texttt{c5.large} & 1$\times$ \texttt{t3.medium} & 1$\times$ \texttt{t3.medium} \\
    Workers            & 2 or 4$\times$ \texttt{c5.xlarge} & 12$\times$ \texttt{c5.2xlarge} & 1$\times$ \texttt{c5.9xlarge} \\
    vCPU per worker    & 4 (2 cores, SMT) & 8 (4 cores, SMT) & 18 (18 cores, SMT off) \\
    RAM per worker     & \SI{8}{\gibi\byte} & \SI{16}{\gibi\byte} & \SI{72}{\gibi\byte} \\
    Processor          & \multicolumn{3}{@{}l@{}}{Xeon Platinum 8124M (A, B), 8223CL (C), 3.0\,GHz} \\
    Allocatable vCPU   & 8 or 16 & 96 & 18 (16 used) \\
    \midrule
    OS                 & \multicolumn{3}{l}{Ubuntu 22.04, kernel 6.8 (aws)} \\
    Kubernetes         & \multicolumn{3}{l}{k3s v1.35.0 (cgroups v2)} \\
    Container runtime  & \multicolumn{3}{l}{containerd 2.1 (bundled with k3s)} \\
    CNI                & \multicolumn{3}{l}{Flannel, VXLAN backend} \\
    Shared storage     & \multicolumn{3}{l}{AWS EFS (NFS, \texttt{ReadWriteMany})} \\
    OpenFOAM           & \multicolumn{3}{l}{v10 (Foundation release)} \\
    MPI                & \multicolumn{3}{l}{Open MPI 4.1, TCP BTL between pods} \\
    \bottomrule
  \end{tabular}
  \par\vspace{3pt}{\footnotesize\raggedright The pitzDaily experiments used a separate two-worker \texttt{t3.xlarge} cluster with the same software stack. CNI = container network interface; BTL = Open~MPI byte transfer layer.\par}
\end{table*}

Worker nodes host the solver pods and each simulation's MPI launcher (50\,m request); the control-plane node hosts no experiment pods. Shared case data reside on AWS EFS, and \texttt{InPlacePodVerticalScaling} is enabled cluster-wide. Worker pods use a custom container image with the software versions in Table~\ref{tab:environment}; the PMPI wrapper is compiled on the cluster and volume-mounted into each pod. MPI launch and isolation follow Section~\ref{sec:k8s-implementation}.

\subsection{Flow Cases and Numerical Settings}
\label{sec:setup_cases}

We use two flow cases, pitzDaily and NACA~0012.  The airfoil case is refined to five mesh resolutions for the different testbeds.  Together, the cases cover two solvers and an approximately 330$\times$ range of mesh size (Table~\ref{tab:cases}).

\begin{table*}[t]
  \centering
  \small
  \caption{Test case summary.}
  \label{tab:cases}
  \begin{tabular}{@{}llclll@{}}
    \toprule
    \textbf{Case} & \textbf{Solver} & \textbf{Cells} & \textbf{Turb.\ model} & \textbf{Ranks} & \textbf{Role} \\
    \midrule
    pitzDaily  & \texttt{simpleFoam}    & 12\,K  & $k$-$\varepsilon$ & 4  & Throttling, overhead (t3) \\
    NACA~0012  & \texttt{rhoSimpleFoam} & 16\,K  & $k$-$\omega$ SST  & 4  & Decomposition vs.\ allocation (A) \\
    NACA~0012  & \texttt{rhoSimpleFoam} & 101\,K & $k$-$\omega$ SST  & 16 & Single-job controls (A) \\
    NACA~0012  & \texttt{rhoSimpleFoam} & 499\,K & $k$-$\omega$ SST  & 16 & Duty cycle, multiplexing (B) \\
    NACA~0012  & \texttt{rhoSimpleFoam} & 1.0\,M & $k$-$\omega$ SST  & 16--22 & Block allocation, core sharing (C) \\
    NACA~0012  & \texttt{rhoSimpleFoam} & 4.0\,M & $k$-$\omega$ SST  & 16--22 & Block allocation at scale (C) \\
    \bottomrule
  \end{tabular}
\end{table*}

\subsubsection{pitzDaily}
\label{sec:case_pitzdaily}

The pitzDaily case is a two-dimensional incompressible backward-facing-step flow with separation, downstream recirculation and reattachment. These spatially distinct flow regions motivate studying how computational work is distributed among subdomains assigned to different MPI ranks.  The mesh contains 12\,225 cells and is solved with \texttt{simpleFoam}, the $k$-$\varepsilon$ model and standard wall functions at $Re_h = 25\,400$.  Supplementary Table~S1 gives the boundary conditions and numerical settings.  This small case is used for the qualitative throttling and network-overhead comparisons on the burstable t3 cluster; its partitions are shown in Figure~\ref{fig:pitzDaily_decomp}.

\subsubsection{NACA~0012 Airfoil}
\label{sec:case_airfoil}

The airfoil case is based on the OpenFOAM tutorial \path{compressible/rhoSimpleFoam/aerofoilNACA0012}.  It solves the two-dimensional steady compressible Reynolds-averaged Navier--Stokes (RANS) equations with \texttt{rhoSimpleFoam} and the $k$-$\omega$ SST model.  All five meshes use the flow and boundary conditions in Table~\ref{tab:naca_bc}.  The C-mesh is generated by \texttt{blockMesh}, with block vertices projected onto a \texttt{triSurfaceMesh} representation of the airfoil, and is one cell thick in the spanwise direction.  Its fine near-wall cells and coarse far field provide the non-uniform workload used in the reservation experiments.

\begin{table}[!t]
  \centering
  \small
  \caption{NACA~0012 flow conditions and numerical settings}
  \label{tab:naca_bc}
  \begin{tabular}{@{}lp{5.4cm}@{}}
    \toprule
    \textbf{Parameter} & \textbf{Value} \\
    \midrule
    Baseline mesh cells & 16\,200 (\texttt{blockMesh\,+\,extrudeMesh}) \\
    Solver           & \texttt{rhoSimpleFoam} (steady compressible) \\
    Thermodynamics   & \texttt{hePsiThermo}, perfect gas, $c_p$ const \\
    Turbulence       & $k$-$\omega$ SST \\
    $U_\infty$       & 250\,m/s; $Ma \approx 0.72$ ($T_\infty=298$\,K) \\
    AoA              & $\alpha = 0^{\circ}$ \\
    $p_\infty$       & $1\times10^{5}$\,Pa \\
    $T_\infty$       & 298\,K \\
    Freestream BC    & \texttt{freestreamVelocity} / \texttt{freestreamPressure} \\
    Wall BC          & No-slip; $k$/$\omega$ wall functions \\
    Pressure limits  & $[0.1\,p_\infty,\;2\,p_\infty]$ \\
    Convergence      & $p{:}\ 10^{-6}$;\; $U,k,\omega,h{:}\ 10^{-5}$ \\
    \bottomrule
  \end{tabular}
  \par\vspace{3pt}{\footnotesize\raggedright The 101\,K, 499\,K, 1.0\,M and 4.0\,M variants keep the flow and boundary conditions and change the mesh resolution; the 4.0\,M variant also changes the pressure-solver agglomeration and the iteration count (Sections~\ref{sec:case_1m4m} and~\ref{sec:setup_stopping}).\par}
\end{table}

The baseline mesh has 16\,200 cells, maximum non-orthogonality below $40^{\circ}$ and maximum skewness below $0.65$.
\phantomsection\label{sec:case_101k}
For the 101\,100-cell mesh, the \texttt{blockMeshDict} counts are increased by approximately 2.5 in each mesh direction: \texttt{zCells}~$60 \to 150$, \texttt{xUCells}~$60 \to 150$, \texttt{xMCells}~$25 \to 62$ and \texttt{xDCells}~$50 \to 125$. Maximum non-orthogonality is $36.5^{\circ}$ and maximum skewness is $0.64$.  These two meshes are used on Testbed~A.

Testbed~B uses 498\,834 hexahedral cells, obtained by further scaling the same tutorial mesh parameters; Supplementary Figure~S1 shows its mesh-generation block topology.  Two auxiliary profiling runs on Testbed~B use 250\,K- and 1.0\,M-cell variants of the same mesh (Figure~\ref{fig:duty-validation}(b)).
\phantomsection\label{sec:case_1m4m}
The Testbed~C meshes contain 997\,578 and 3\,990\,312 cells. Relative to the 499\,K mesh, their cell counts in each in-plane direction are scaled by $\sqrt{2}$ and $2\sqrt{2}$, respectively, doubling and octupling the total cell count.  On the 4.0\,M mesh, the GAMG coarse levels are agglomerated across processors with \texttt{processorAgglomerator masterCoarsest}; this reduces but does not remove the dependence of pressure-solver work on the partition. No field is written during a timed run on this mesh.  The decomposition and resource configurations are defined in Section~\ref{sec:setup_configs}, and the stopping rules in Section~\ref{sec:setup_stopping}.

\FloatBarrier
\subsection{Experimental Design}
\label{sec:setup_configs}

\begin{table*}[!b]
  \centering
  \small
  \caption{Overview of the experimental comparisons}
  \label{tab:comparisons}
  \begin{tabular}{@{}p{4.2cm}p{2.4cm}p{7.6cm}l@{}}
    \toprule
    \textbf{Comparison} & \textbf{Testbed} & \textbf{Compared settings} & \textbf{Results} \\
    \midrule
    Single-simulation CPU settings & A & decomposition, request vectors and CPU limits & \ref{sec:res_efficiency}--\ref{sec:res_scaling} \\
    CPU throttling & t3 (pitzDaily) & identical requests with and without limits & \ref{sec:res_throttle} \\
    Co-location interference & B & one or two simulations; request policies & \ref{sec:res_concurrent} \\
    Concurrency scaling & B & one to eight simulations; request policies and MPI waiting modes & \ref{sec:res_duty} \\
    In-place CPU adjustment & B & separate-pass and online profiling; static throughput reference & \ref{sec:res_dynamic} \\
    Rank allocation & C & zone-based layouts, balanced Scotch and core-sharing variants & \ref{sec:res_blocks} \\
    \bottomrule
  \end{tabular}
\end{table*}

Single-simulation controls separate the effects of decomposition, CPU limits, reservation total and request shape. Co-located runs measure admission, interference and throughput as concurrency increases. Controller runs evaluate request adjustment during solver execution, while a separate fixed-core comparison examines rank allocation and core sharing. Table~\ref{tab:comparisons} summarizes these comparisons and their corresponding results.
\FloatBarrier

\subsubsection{Single-Simulation CPU Controls}
\label{sec:setup_single}

\begin{figure}[h]
  \centering
  \includegraphics[width=\columnwidth]{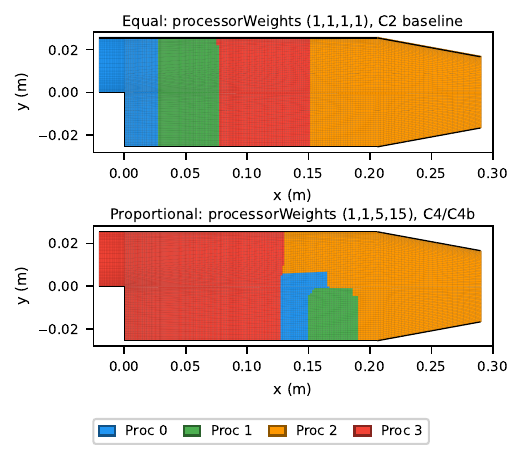}
  \caption{pitzDaily mesh colored by Scotch subdomain. \emph{Top}: equal decomposition (C2 baseline), all four subdomains $\approx$3\,056 cells each. \emph{Bottom}: proportional decomposition, processorWeights $(1,1,5,15)$ gives 556\,/\,556\,/\,2\,778\,/\,8\,335 cells.  Proc~3 (red, 8\,335 cells) covers the inlet channel, the step, and the upstream half of the expanded channel ($x < 0.13$\,m); Proc~0 and Proc~1 (blue, green; 556 cells each) are two small blocks on the lower wall between $x = 0.13$ and 0.19\,m, and Proc~2 (orange) covers the rest of the downstream channel.  Flow is from left to right.}
  \label{fig:pitzDaily_decomp}
\end{figure}

\begin{figure*}[t]
  \centering
  \includegraphics[width=\textwidth]{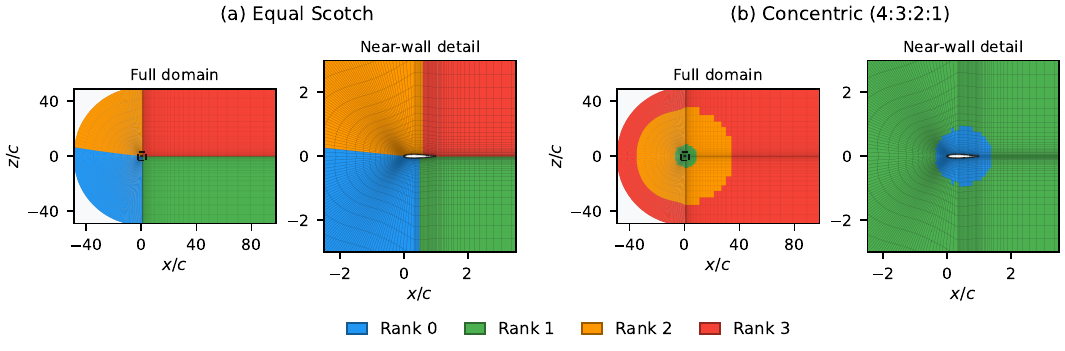}
  \caption{Four-rank partitions of the 16\,K-cell NACA~0012 mesh on Testbed~A. (a)~Equal Scotch decomposition (C2 baseline), with approximately 4\,050 cells per rank. (b)~Distance-based concentric decomposition with cell-count ratio 4:3:2:1. Each panel pairs the full domain with a near-wall detail; dashed boxes mark the enlarged region. Colors identify individual ranks. Rank~0 is the innermost concentric zone (6\,480 cells), and Rank~3 is the outermost (1\,620 cells).}
  \label{fig:airfoil_decomp_prop}
\end{figure*}

\begin{table*}[t]
  \centering
  \small
  \caption{16-rank configurations on Testbed~A}
  \label{tab:configs16}
  \begin{tabular}{@{}llllc@{}}
    \toprule
    \textbf{ID} & \textbf{Decomposition} & \textbf{CPU per rank} & \textbf{CPU limits} & \textbf{Total (vCPU)} \\
    \midrule
    C2-16  & Scotch equal & 1000\,m $\times$ 16, limits = requests & yes & 16.0 \\
    C2-16b & Scotch equal & 1000\,m $\times$ 16, requests only   & no  & 16.0 \\
    C2-16w & Scotch W     & 1000\,m $\times$ 16, requests only   & no  & 16.0 \\
    C4-16  & Scotch W     & 182\,/\,909\,/\,2727\,m, requests only & no & 16.0 \\
    C4-16L & Scotch W     & 67\,/\,333\,/\,1000\,m, requests only  & no & 5.9 \\
    C4-16U & Scotch W     & 367\,m $\times$ 16, requests only      & no & 5.9 \\
    \bottomrule
  \end{tabular}
  \par\vspace{3pt}{\footnotesize\raggedright Four c5.xlarge workers, four ranks per node. W = Scotch with grouped \texttt{processorWeights} $(1,1,5,15)\times4$. Variants of one configuration: b removes the CPU limits of C2-16; w changes C2-16b to weighted Scotch while retaining equal requests and no CPU limits; L reduces the proportional requests of C4-16 to a 5.9\,vCPU total; U gives every rank the same reduced request at that total. Totals are nominal (vCPU); the deployed 16.0\,vCPU manifests sum to 15.94\,vCPU (Section~\ref{sec:setup_configs}).\par}
\end{table*}

Single-simulation configurations use indices C0 to C5; these encode neither rank count nor simulation count. The suffix \mbox{-16} denotes a 16-rank variant, with further letters distinguishing configurations (Table~\ref{tab:configs16}). The four-rank experiments use pitzDaily and the 16\,K NACA mesh (Supplementary Table~S2). Their baseline C2 combines equal Scotch decomposition with equal CPU requests.

For pitzDaily, equal Scotch decomposition gives approximately equal cell counts.  Weighted Scotch with $(1,1,5,15)$ produces subdomains of 556, 556, 2\,778 and 8\,335 cells (Figure~\ref{fig:pitzDaily_decomp}). The weights specify target cell shares, and Scotch determines the spatial boundaries.  Under \texttt{simpleFoam} with $k$-$\varepsilon$, each cell performs the same operations per SIMPLE iteration; wall functions act at boundary faces, not as additional volume operations.  Cell count is therefore an approximate per-iteration cost proxy, with Proc~3 carrying about 68\% of the work.

\phantomsection\label{sec:case_airfoil_decomp}
On the 16\,K NACA mesh, equal Scotch produces four sectors of approximately 4\,050 cells each.  The manual alternative ranks cells by Euclidean distance from the airfoil midchord $(0.5c,\,0)$ and assigns them to four concentric zones with cell-count ratio 4:3:2:1. Proc~0 is the innermost zone with 6\,480 cells and Proc~3 the outermost with 1\,620 cells (Figure~\ref{fig:airfoil_decomp_prop}).  This manual assignment fixes radial membership, whereas Scotch chooses the subdomain geometry through graph partitioning.

C3 applies requests and hard limits of 0.25/\allowbreak0.25/\allowbreak1.0/\allowbreak2.5\,vCPU, using the fractions from $(1,1,5,15)$ in Eq.~\eqref{eq:cpu_alloc}, rounded up to 250\,m with the dense rank taking the remainder of the 4.0\,vCPU budget. C4 uses the same requests without limits, and C4b uses the exact fractions, 182/182/909/2727\,m, without limits.  For NACA, C2-manual holds requests equal while changing to the concentric partition; C4~manual then applies the pitzDaily request vector to that partition. The vector is not matched to the NACA cell-count ratio $(4,3,2,1)$: Proc~0 receives 250\,m despite holding 6\,480 cells.  This pair separates the decomposition effect from the effect of the inherited CPU vector (Section~\ref{sec:res_efficiency}).  C5 is the separate preset-phase resize configuration described in Section~\ref{sec:setup_controller}.

Network-path controls accompany these four-rank CPU comparisons. C0 runs four MPI processes in two pods with \texttt{hostNetwork:\ true}, bypassing Flannel VXLAN; C2 uses four pods on the pod network. C1 runs all four processes in one pod with shared memory and provides a single-node lower bound rather than a direct comparator for cross-node configurations.

At 16 ranks on the 101\,K mesh, the decompositions are equal Scotch and Scotch with grouped \texttt{processorWeights} $(1,1,1,1,\;1,1,1,1,\;5,5,5,5,\;15,15,15,15)$. The four groups contain four ranks each, with target cell fractions $1/88$, $1/88$, $5/88$ and $15/88$ per rank.  These weights prescribe cell shares; they do not prescribe manual concentric zones. Table~\ref{tab:configs16} gives the CPU settings.  C2-16 and C4-16 change decomposition, limits and requests together, so the following controls separate their effects.  C2-16b removes the CPU limits from C2-16.  C2-16w changes to weighted Scotch while retaining equal 1000\,m requests, making its comparison with C4-16 a test of the request shape.  C4-16L retains that shape but reduces the requests to $67/333/1000$\,m, lowering the nominal total from 16.0 to 5.9\,vCPU.  C4-16U assigns 367\,m to every rank at the same reduced total and was run in a second session with C4-16L and C2-16w as same-session anchors.

Placement is identical across these 16-rank configurations: each of the four workers hosts one rank from each weight group.  The pods have no memory limits, so C2-16 is Burstable with a CPU quota, not Guaranteed.  In the 16.0\,vCPU configurations, rank~0 requests 60\,m less than nominal to accommodate the 50\,m launcher on a worker; the deployed C4-16 total is thus 15\,940\,m rather than 16\,000\,m.  The 5.9\,vCPU configurations leave room for the launcher without this adjustment.

\FloatBarrier
\subsubsection{CPU Multiplexing and Contention}
\label{sec:setup_colocation}

\begin{figure*}[t]
  \centering
  \includegraphics[width=0.82\textwidth]{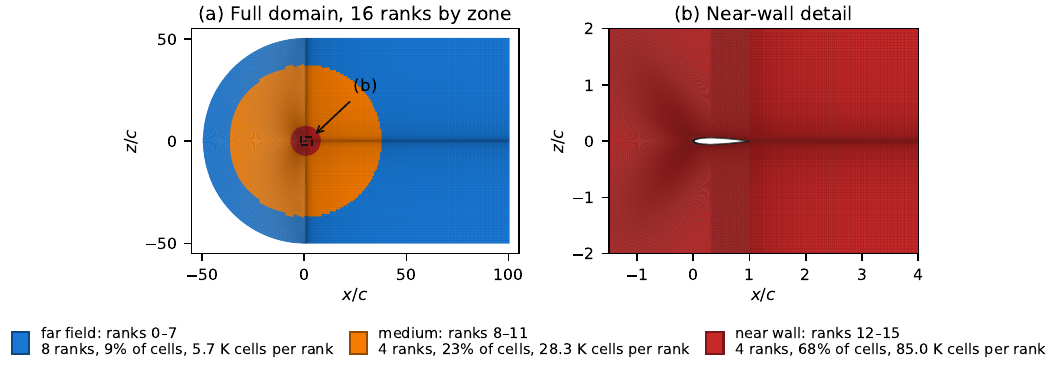}
  \caption{NACA~0012 mesh of the Testbed~B co-location and controller experiments (498\,834 cells, 16 ranks), colored by the three zones of the concentric partition. (a)~Full domain; the dashed box marks the near-wall detail shown in (b). The legend gives each zone's ranks, its share of the cells, and its average cells per rank.}
  \label{fig:decomposition500k}
\end{figure*}

\begin{table}[t]
  \centering
  \small
  \caption{Multiplexing configurations on Testbed~B}
  \label{tab:configs_mux}
  \setlength{\tabcolsep}{2.8pt}
  \begin{tabular}{@{}llccc@{}}
    \toprule
    \textbf{Config} & \textbf{Sims} & \textbf{Pods} &
    \textbf{CPU/rank} & \textbf{Reserved} \\
    \midrule
    C-1E & 1 & 16 & 1000\,m (equal)    & 16.0\,vCPU \\
    C-1P & 1 & 16 & 67--1005\,m (prop.) & 5.9\,vCPU \\
    C-2E & 2 & 32 & 1000\,m (equal)    & 32.0\,vCPU \\
    C-2P & 2 & 32 & 67--1005\,m (prop.) & 11.8\,vCPU \\
    C-$N$P & 3--8 & $16N$ & 67--1005\,m (prop.) & $5.9N$\,vCPU \\
    C-$N$E & 3--6 & $16N$ & 1000\,m (equal) & $16N$\,vCPU \\
    C-$N$P-yield & 6--8 & $16N$ & 67--1005\,m, yield & $5.9N$\,vCPU \\
    C-$N$U & 1, 2, 6 & $16N$ & 369\,m (uniform) & $5.9N$\,vCPU \\
    C-$N$U-yield & 7, 8 & $16N$ & 369\,m, yield & $5.9N$\,vCPU \\
    \bottomrule
  \end{tabular}
  \par\vspace{3pt}{\footnotesize\raggedright Twelve c5.2xlarge workers. C-$N$X: $N$ concurrent simulations under request policy X, with E = 1000\,m per rank, P = the proportional vector 67/335/1005\,m and U = 369\,m per rank at the same nominal total as P. CPU values are per-rank requests; no limits are set. The -yield variants launch with yielding MPI waits (\texttt{mpi\_yield\_when\_idle}).\par}
\end{table}

\phantomsection\label{sec:case_500k}
The co-location and controller experiments on Testbed~B use the same 499\,K mesh and 16-rank manual partition.  Cells are grouped by distance from the airfoil center ($x/c = 0.5$) into three concentric zones and then split into equal angular sectors within each zone.  With weights $(1,1,1,1,1,1,1,1,\;5,5,5,5,\;15,15,15,15)$, the four dense ranks (12--15, weight~15) hold the innermost 68\% of cells, the four medium ranks (8--11, weight~5) the next 23\%, and the eight sparse ranks (0--7, weight~1) the outermost 9\%.  Figure~\ref{fig:decomposition500k} shows the resulting load regions and their average cell counts per rank.

With this partition fixed, the initial $2 \times 2$ design crosses equal or proportional requests with one or two simulations. Configurations are written C-$N$X: $N$ is the number of concurrent simulations and X is the request policy, equal (E), proportional (P) or uniform (U). Every configuration uses requests without limits (Table~\ref{tab:configs_mux}). Equal requests reserve 16.0\,vCPU per simulation. The proportional vector is $67/335/1005$\,m, from Eq.~\eqref{eq:cpu_alloc} at $C_{\text{job}} = 5.9$\,vCPU; Testbed~A instead uses $67/333/1000$\,m, totaling 5.87\,vCPU. C-2E thus reserves 32\,vCPU, one third of the cluster, and C-2P reserves 11.8\,vCPU.

The proportional sweep extends to $N = 3, \ldots, 8$.  Nominal reservation increases by 5.9\,vCPU per simulation, from 17.7\,vCPU at $N=3$ to 47.2\,vCPU at $N=8$, within the 96-vCPU allocatable capacity.  Equal-request controls cover $N = 3$--$6$ with 1000\,m per rank; uniform controls C-$N$U use 369\,m per rank at the same nominal 5.9-vCPU total as the proportional vector.  Each uniform configuration is paired with a proportional anchor in the same session.  Waiting-mode comparisons enable \texttt{mpi\_yield\_when\_idle} for proportional requests at $N = 6$--$8$ and for the uniform comparisons at $N = 7$ and $8$.

\FloatBarrier
\subsubsection{In-Place CPU Request Adjustment}
\label{sec:setup_controller}

The two end-to-end controller runs use the 499\,K NACA case on Testbed~B.  Both start with one simulation at equal requests and admit up to six.  Their load signals come from either a separate 50-iteration profiling pass or a PMPI window read after 240\,s of production execution (Section~\ref{sec:controller-design}).  Static six-way co-location provides a throughput reference for the cost of serial admission.

C5 checks the resize mechanism separately with four ranks on the burstable cluster.  It extends C4 with preset Phase~1 requests of 500/500/1200/1800\,m and Phase~2 requests of 1000\,m per rank, applied at iteration~100 through \texttt{kubectl patch -{}-subresource resize}. This protocol is distinct from the measurement-driven controller (Section~\ref{sec:meth_phase_scaling}).

\subsubsection{Rank Allocation at a Fixed Core Budget}
\label{sec:meth_blocks}

The Testbed~C comparison varies the partition under a fixed budget of 16 physical cores.  Supplementary Figure~S2 illustrates the rank layouts. A request cannot give a single-threaded rank more than one CPU of simultaneous execution (Section~\ref{sec:meth_request_semantics}); assigning more cores to an expensive zone therefore requires more ranks in that zone.

Layouts are written as far-field/medium/near-wall rank counts. For the one-rank-per-core comparisons, the three zones retain their cell shares of 9, 23 and 68\%, while 8/4/4, 4/4/8, 2/4/10 and 2/2/12 progressively allocate more of the sixteen ranks to the near-wall zone.  The 8/4/4 layout is the concentric reference used on Testbed~B (Section~\ref{sec:case_500k}); 2/4/10 assigns ten cores to the near-wall zone.  The four layouts were selected before the runs, without an optimizer, and balanced Scotch provides an equal-cell reference.

The core-sharing variants retain either 8/4/4 or 8/4/10 and place two or four far-field ranks on a shared core.  In 8/4/10, the eight far-field ranks share two cores, four per core, so 22 ranks occupy 16 cores.  Pinning enforces the sharing, and sub-core requests set reservation and cgroup CPU weights (Section~\ref{sec:setup_hw}).  The one-rank-per-core layouts use busy-waiting; the sharing variants use yielding waits, except for one busy-wait run that was stopped. Every rank requests 1000\,m except the shared ranks, which request 500 or 250\,m; no limits are set.  One unpinned variant uses 250\,m for the shared ranks and leaves placement to the Linux fair scheduler. Table~\ref{tab:sharing} identifies the variants and reports their results.  Short and extended runs use the stopping rules in Section~\ref{sec:setup_stopping}; their wall-clock and pressure-iteration metrics are defined in Section~\ref{sec:setup_measurements}.

\FloatBarrier
\subsection{Measurements and Reproducibility}
\label{sec:setup_metrics}

\subsubsection{Stopping Criteria}
\label{sec:setup_stopping}

The pitzDaily runs stop when residuals fall below $10^{-5}$ (Supplementary Table~S1).  The 16\,K and 101\,K NACA runs use the residual thresholds in Table~\ref{tab:naca_bc}, with a cap of 5\,000 iterations.  Testbed~B uses 200 iterations on the 499\,K mesh for timing and PMPI profiling.  The short Testbed~C runs use 200 iterations on the 1.0\,M mesh and 50 on the 4.0\,M mesh, where each iteration takes 5--13\,s.

For the extended Testbed~C comparison, one run each of 8/4/4, 2/4/10 and Scotch uses the prescribed residual criterion with a cap of 4000 iterations on the 1.0\,M mesh, and a fixed 400 iterations on the 4.0\,M mesh.  None of the 4000-iteration runs reaches the criterion, and all retain residual oscillations of similar magnitude (Section~\ref{sec:res_blocks}).  These are fixed-iteration comparisons, not measurements of time to a converged result.

\subsubsection{Timing and Resource Measurements}
\label{sec:setup_measurements}

Single-simulation wall-clock time ($T$) is the solver's \texttt{ClockTime}, measured to the applicable stopping condition. \texttt{ExecutionTime} is CPU time and is reported only where the two differ.  For concurrently started simulations, $T_N$ is the makespan, the completion time of the slowest simulation, and $T_1$ is the corresponding single-simulation reference.  Relative throughput is $\Theta(N) = N\,T_1 / T_N$; degradation of the slowest simulation is $(T_N - T_1)/T_1$, and run-level fairness is the ratio of the slowest to the fastest simulation.

For the controller, pipeline time runs from the first profiling start (pass-based design) or first production start (online design) to the last completion, excluding initial deployment.  Batch throughput is $N\,T_1/T_{\text{pipeline}}$, using C-1P as the reference ($T_1 = 1249$\,s).  The run-level fairness measure differs from the controller's fairness check, which compares iteration rates over one interval (Section~\ref{sec:controller-design}).

CPU utilization ($U$) is sampled per pod with \texttt{kubectl top pod} at \SI{5}{\second} intervals and from cgroup \texttt{cpu.stat}; node CPU utilization and per-pod memory are sampled at the same interval from the Kubernetes metrics API. Per-rank duty cycle $\bar{d}_i$ comes from PMPI (Eq.~\eqref{eq:duty-cycle}), while per-pod occupancy comes from cgroup accounting.  Throttling is measured by \texttt{nr\_throttled} and \texttt{throttled\_usec} in \path{/sys/fs/cgroup/cpu.stat}.  Reserved CPU time ($R_{\text{CPU}}$, vCPU-seconds) is the job's total CPU request multiplied by its wall-clock time when that total is constant.

\phantomsection\label{sec:setup_pressure_cost}
For Testbed~C, pressure-iteration cost is wall-clock time divided by the total number of pressure-solver (GAMG) iterations.  It includes the momentum, turbulence, and transport work amortized over those pressure iterations.  Together with the pressure-iteration count, it decomposes the measured wall-clock ratio into cost per unit of solver work and the number of units required; the two factors are not independent measurements.

\needspace{6\baselineskip}
\subsubsection{Repeated Runs and Numerical Consistency}
\label{sec:setup_repeats}

Timed configurations on Testbeds~A and~B use three runs unless specified otherwise.  Repeated timings and error bars use the mean $\pm$ population standard deviation.  On Testbed~B, C-3P and C-4P use six runs, and C-8P, whose runs take 5.5~hours each, uses two.  Detailed counts for the same-session uniform and yielding comparisons accompany Tables~\ref{tab:uniform_mux} and~\ref{tab:yield}, including which comparisons have only one run.  Profiling and end-to-end controller evaluations are single runs.  Testbed~C uses three runs on the 1.0\,M mesh and two on the 4.0\,M mesh; the 4000- and 400-iteration runs are single runs.  The pitzDaily results are single-run observations and are used qualitatively.

Within-session variability is summarized in the Supplementary Material. With three runs, differences of a few percent are treated as observations, not robust performance advantages.

The Testbed~B campaign contains 274 completed simulation executions; their breakdown by configuration and controller phase is given in the Supplementary Material. Five additional profiling runs, excluded from this total, are reported in Section~\ref{sec:res_waiting_profile}. Each simulation in a concurrent run counts as one execution, but the run is the statistical unit because its simulations share one condition. Two yielding runs at $N=7$ were lost to a worker-node failure before completion; they were excluded and repeated.

Numerical consistency is checked by comparing residual histories and stopping iteration counts across request settings within each partition. The matching histories and iteration counts are reported in the Supplementary Material and with the timing comparisons.

\section{Results and Discussion}
\label{sec:results}

Lower CPU reservations are useful for shared-cluster execution if they preserve acceptable single-simulation performance and support co-location within the cluster's CPU capacity. Section~\ref{sec:res_single} first separates the effects of decomposition, limits, and requests on one simulation. Section~\ref{sec:res_duty} then measures the reservation, throughput, and contention of co-located simulations, and Section~\ref{sec:res_dynamic} examines online profiling and in-place adjustment of running solvers. The fixed-core rank allocation comparison in Section~\ref{sec:res_blocks} evaluates a separate partition-side choice. Section~\ref{sec:discussion} discusses the implications and scope of these results.

\subsection{CPU Reservation and Single-Job Performance}
\label{sec:res_single}

The single-job controls distinguish changes to the work assigned to each rank from changes to its CPU reservation and execution constraints. Testbed~A provides the quantitative comparisons at 4 and 16 ranks; the pitzDaily runs illustrate the effect of sub-core limits. The section ends with the cost of lowering the total reservation while keeping the rank allocation fixed.

\subsubsection{Decomposition and CPU Request Effects}
\label{sec:res_efficiency}

\begin{figure*}[t]
  \centering
  \includegraphics[width=\textwidth]{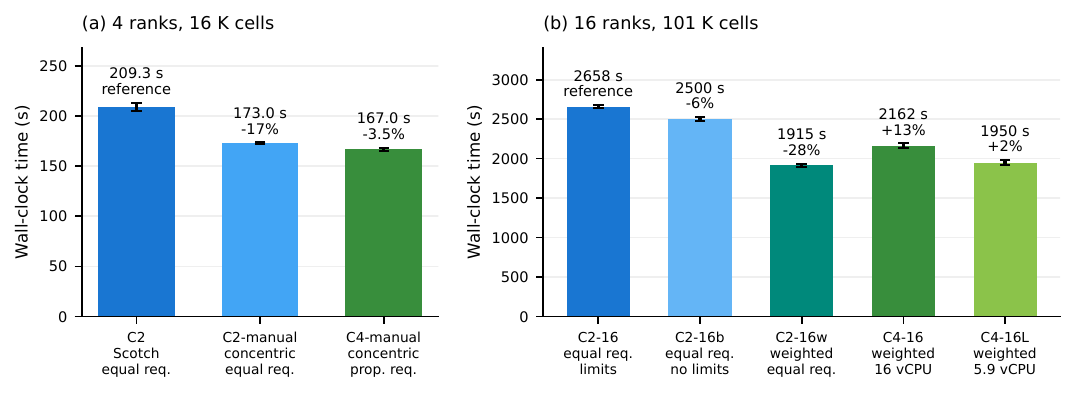}
  \caption{NACA~0012 single-job controls on Testbed~A (mean $\pm$ population standard deviation, 3 runs per configuration). (a)~4 ranks, 16\,K cells: the $-3.5\%$ annotation compares C4-manual with C2-manual; $-17\%$ compares C2-manual with C2. (b)~16 ranks, 101\,K cells: C2-16b and C2-16w are compared with C2-16; C4-16 and C4-16L with C2-16w. Each panel uses same-session wall-clock measurements; the 16-rank iteration counts are in Table~\ref{tab:naca_16rank}.}
  \label{fig:naca_c5}
\end{figure*}

\begin{table*}[t]
  \centering
  \small
  \caption{NACA~0012 wall-clock times at 16 ranks (101\,K cells, Testbed~A)}
  \label{tab:naca_16rank}
  \begin{tabular}{@{}llrrrrr@{}}
    \toprule
    \textbf{Config} & \textbf{Change from its reference} & \textbf{Wall clock (s)} &
      \textbf{Iter.} & \textbf{s/iter} & \textbf{vs C2-16} & \textbf{vs C2-16w} \\
    \midrule
    C2-16  & reference: Scotch equal, 1000\,m with limits & $2658 \pm 21$ & 1390 & 1.912 & baseline & -- \\
    C2-16b & C2-16 without CPU limits            & $2500 \pm 24$ & 1390 & 1.798 & $-6\%$ & -- \\
    \textbf{C2-16w} & \textbf{C2-16b with the weighted decomposition} & $\mathbf{1915 \pm 21}$ & 1449 & 1.322 & $\mathbf{-28\%}$ & baseline \\
    C4-16  & C2-16w with requests $182/909/2727$\,m & $2162 \pm 32$ & 1449 & 1.492 & $-19\%$ & $+13\%$ \\
    C4-16L & C2-16w with requests $67/333/1000$\,m (5.9\,vCPU) & $1950 \pm 31$ & 1449 & 1.346 & $-27\%$ & $+2\%$ \\
    \bottomrule
  \end{tabular}
  \par\vspace{3pt}{\footnotesize\raggedright 3 runs each, one session. Iter.\ = iterations to convergence. Configurations are defined in Table~\ref{tab:configs16}; times are wall-clock (ClockTime).\par}
\end{table*}

At four ranks, the concentric NACA~0012 decomposition with equal requests (C2-manual) completes in 173.0\,s, 17\% faster than the equal Scotch partition (C2, 209.3\,s; Figure~\ref{fig:naca_c5}(a) and Supplementary Table~S3). Its more regular inter-subdomain boundaries accompany the lower time, but the measurements do not separate communication volume from convergence behavior. Applying the $(1,1,5,15)$ request vector to this partition gives 167.0\,s, a further 3.5\% reduction relative to C2-manual. The vector is inverted relative to the 4:3:2:1 cell counts: the heaviest rank receives the smallest request. With two ranks on two physical cores per node, the vector has little CPU contention to arbitrate. Decomposition accounts for most of the observed difference.

At 16 ranks, the controls in Table~\ref{tab:naca_16rank} separate three changes that distinguish C2-16 from C4-16: decomposition, CPU limits, and request vector. Removing the limits reduces wall-clock time by 6\% (C2-16b against C2-16). Changing to the weighted decomposition while retaining equal requests reduces it by a further 23\% (C2-16w against C2-16b), or 28\% relative to C2-16. Time per iteration falls by 26.5\%, from 1.798\,s to 1.322\,s, while the weighted partition requires 4\% more iterations to meet the same convergence criteria. Thus, the main improvement in this comparison comes from the decomposition, even after accounting for the different iteration counts.

On that same weighted partition, the proportional 16.0\,vCPU vector C4-16 is 13\% slower than equal requests. Four ranks share two physical cores per worker on this SMT testbed, so weights can affect service when runnable ranks contend. The result is consistent with a low-weight sparse rank being delayed during its short compute phase, but scheduling traces were not collected to establish that mechanism. These comparisons use one provisioning session; the request-vector effects vary between sessions, as the reduced-budget controls below show.

\subsubsection{CPU Limits and MPI Synchronization}
\label{sec:res_throttle}

\begin{figure*}[t]
  \centering
  \includegraphics[width=\textwidth]{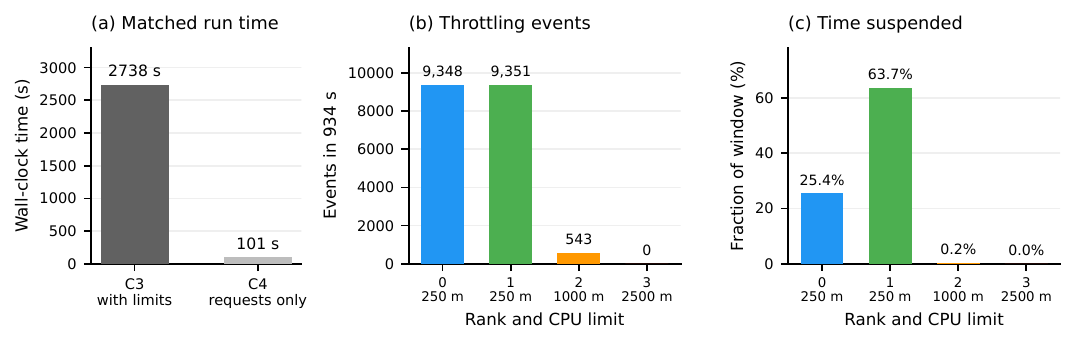}
  \caption{CPU limits and MPI execution in pitzDaily. (a)~C3 and C4 have the same decomposition and requests; only C3 sets limits equal to those requests. (b)~Per-rank increments of \texttt{nr\_throttled} and (c)~fraction of time suspended, sampled at 6.6\,s intervals over the same 934\,s window of C3. C4 records zero throttling on all ranks. These are single-run observations on burstable t3.xlarge instances; the wall-clock ratio is illustrative.}
  \label{fig:throttling}
\end{figure*}

The pitzDaily comparison shows why the low-request configurations are deployed without CPU limits. C3 and C4 use the same decomposition and requests, $(0.25,0.25,1.0,2.5)$\,vCPU; C3 also sets limits equal to these requests. Its wall-clock time is 2738\,s against C4's 101\,s, about 27$\times$ longer (Figure~\ref{fig:throttling}(a)). The 78$\times$ gap to the equal configuration C2 also includes a decomposition change. These are single runs on burstable instances, so the ratios indicate the scale of the effect rather than precise timing estimates.

The per-rank bandwidth counters provide the mechanism evidence. During a 934\,s sampling window, the two ranks limited to 250\,m were throttled in 9348 and 9351 of their 9351 CFS periods and spent 25\% and 64\% of wall-clock time suspended. The 1000\,m rank was throttled in 6\% of periods for 0.2\% of the window, while the 2500\,m rank was never throttled. A 250\,m limit allows 25\,ms of CPU time per 100\,ms period, whereas a single-threaded rank's instantaneous demand can reach approximately 1000\,m. Repeated suspension delays the rank's progress and consequently the collectives that require it. With requests only, all four C4 throttle counters remain zero.

This result concerns limits equal to sub-core requests. It does not rule out limits at or above a rank's peak demand, which were not tested for the reduced-request configurations. The equal baselines retain their 1000\,m limits, whose smaller cost is measured by C2-16b. Supplementary Figure~S3 shows per-rank CPU utilization for C2 and C4, and Figure~S4 reports wall-clock times for all seven pitzDaily configurations. The communication-path checks are described in Supplementary Section~S1; they do not isolate the pod overlay's cost.

\subsubsection{Reducing Reservations at a Fixed Rank Allocation}
\label{sec:res_scaling}

\begin{table}[h]
  \centering
  \small
  \caption{Uniform against proportional low requests at 16 ranks (Testbed~A)}
  \label{tab:naca_16rank_uniform}
  \footnotesize
  \setlength{\tabcolsep}{3.5pt}
  \begin{tabular}{@{}llrrr@{}}
    \toprule
    \textbf{Config} & \textbf{Requests} & \textbf{Reserved} & \textbf{Time (s)} & \textbf{vs C2-16w} \\
    \midrule
    C2-16w & $1000$\,m $\times$ 16 & 16.0\,vCPU & $1783 \pm 26$ & baseline \\
    C4-16L & $67/333/1000$\,m      & 5.9\,vCPU  & $1531 \pm 8$  & $-14\%$ \\
    C4-16U & $367$\,m $\times$ 16  & 5.9\,vCPU  & $1564 \pm 9$  & $-12\%$ \\
    \bottomrule
  \end{tabular}
  \par\vspace{3pt}{\footnotesize\raggedright 101\,K cells, 3 runs each, one session on 6 September 2026, wall-clock times. All three configurations use the weighted Scotch decomposition and requests only.\par}
\end{table}

Lowering the total request from 16.0 to 5.9\,vCPU reduces the reservation by 63\% while keeping the 16-rank weighted partition fixed. C4-16L completes in 1950\,s, within 2\% of the equal-request C2-16w and 10\% faster than the proportional 16.0\,vCPU configuration C4-16 (Table~\ref{tab:naca_16rank}). C4-16 itself releases no cluster-level reservation: the 82\% reduction for each far-field rank is offset by larger near-wall requests. The lower job total in C4-16L is what creates reservation headroom.

The reduction does not require a proportional shape. A second session compares $67/333/1000$\,m with a uniform 367\,m per rank, both totaling about 5.9\,vCPU, alongside a same-session equal-request anchor (Table~\ref{tab:naca_16rank_uniform}). The uniform vector takes 1564\,s against the proportional vector's 1531\,s, a 2\% difference; both are 12--14\% faster than equal requests in this session. In the earlier session, the proportional low-request vector was 1.8\% slower than equal requests. The observed cost of reducing the reservation on Testbed~A therefore never exceeded 2\%, but its direction changed between sessions. Section~\ref{sec:res_concurrent} measures the corresponding single-job cost and co-location benefit on Testbed~B.

\FloatBarrier
\subsection{CPU Multiplexing Through Co-location}
\label{sec:res_duty}

Testbed~B keeps the mesh and rank partition fixed while varying requests and the number of co-located simulations. The results connect three quantities: the reservation entered in the scheduler's accounting, the throughput delivered by the shared workers, and the number of runnable ranks competing for their CPUs.

\subsubsection{MPI Duty Cycles and CPU Occupancy}
\label{sec:res_waiting_profile}

\begin{figure*}[t]
  \centering
  \includegraphics[width=\textwidth]{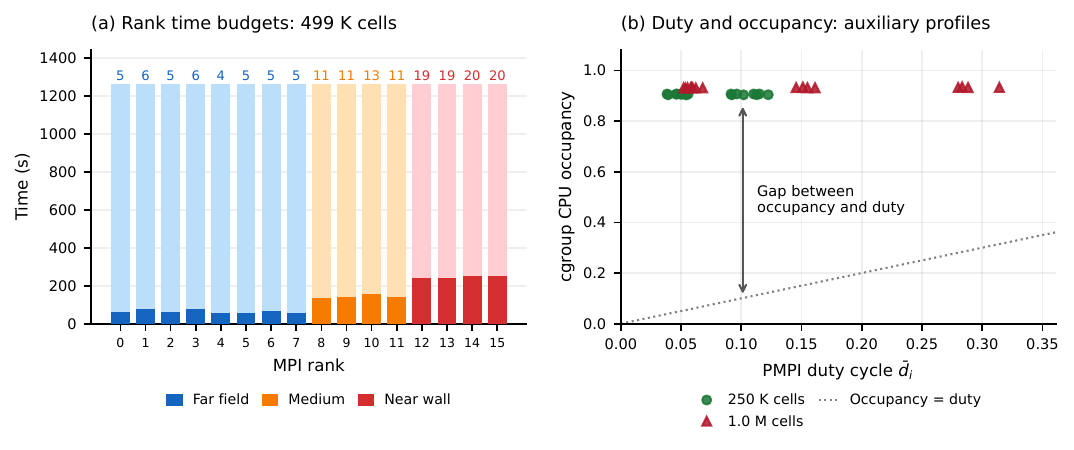}
  \caption{MPI waiting and CPU occupancy on Testbed~B. (a)~The 499\,K-cell main case, 16 ranks and 200 iterations: time outside intercepted MPI calls (solid, an upper bound on computation) and inside them (lighter shade); numbers above the bars are duty cycles in percent. (b)~Separate auxiliary profiles on the 250\,K- and 1.0\,M-cell variants: PMPI duty cycle against whole-run cgroup CPU occupancy from \texttt{cpu.stat}. The diagonal denotes equality. High occupancy despite low duty is consistent with busy-polling MPI waits.}
  \label{fig:duty-cycle}
  \label{fig:duty-validation}
\end{figure*}

The 499\,K-cell, 16-rank profile shows substantial MPI waiting throughout the partition (Figure~\ref{fig:duty-cycle}(a)). Far-field ranks average a 5.0\% duty cycle, spending 95\% of their time inside the intercepted MPI calls; medium and near-wall ranks average 11.5\% and 19.4\%, respectively. Even the densest ranks spend 80\% of the run in MPI. The reservation-weighted waiting fraction of Eq.~\eqref{eq:reclaimable} is 90\% under equal 16\,vCPU requests and 83\% under the proportional 5.9\,vCPU vector. These fractions describe the reservation associated with MPI waiting, not CPU cycles already available to another rank.

High MPI waiting also occurs with equal Scotch partitioning on the same testbed: duty cycles range from 7.7 to 13.1\% (87--92\% waiting), and wall-clock time is within 1\% of the concentric result (1272\,s against 1263\,s). The concentric partition concentrates the waiting in identifiable rank groups. Unlike the Testbed~A comparison, this profile shows little wall-clock effect of partitioning; rank density, mesh size and placement differ between the testbeds, and their contributions are not isolated.

The rank-group pattern also persists with \texttt{hostNetwork}. Sparse-rank duty cycles are 4.7--5.8\% with host networking against 4.4--6.0\% with pod networking; dense-rank values are 19.8--22.7\% against 18.9--19.9\%. This configuration check retains the sparse-to-dense contrast under both network paths.

In the auxiliary 250\,K- and 1.0\,M-cell profiles, CPU occupancy approaches a full logical CPU even for ranks whose PMPI duty cycle is below 10\% (Figure~\ref{fig:duty-validation}(b)). The gap is consistent with the MPI progress engine busy-polling while waiting. CPU-usage counters alone therefore cannot distinguish solver computation from MPI waiting, motivating the PMPI load signal used by the controller. A waiting rank also remains runnable, so admission must account for worker counts as well as reservations (Eq.~\eqref{eq:capacity-boundary}).

\subsubsection{Reservation, Admission, and Throughput}
\label{sec:res_concurrent}
\label{sec:res_pareto}

\begin{table*}[t]
  \centering
  \caption{Concurrent simulation results on Testbed~B}
  \label{tab:results_mux}
  \small
  \begin{tabular}{@{}lrrrrr@{}}
    \toprule
    \textbf{Config} & \textbf{Sim\,A (s)} & \textbf{Sim\,B (s)} &
    \textbf{Makespan (s)} & \textbf{Thpt.} & $T_A/T_B$ \\
    \midrule
    C-1E & $1179 \pm 23$ & -- & 1179 & 1.00$\times$ & -- \\
    C-1P & $1249 \pm 17$ & -- & 1249 & 1.00$\times$ & -- \\
    C-2P & $1334 \pm 10$ & $1412 \pm 9$ & $1412 \pm 9$ & \textbf{1.77$\times$} & 0.94 \\
    C-2E & $1289 \pm 11$ & $1277 \pm 18$ & $1289 \pm 11$ & 1.83$\times$ & 1.01 \\
    \bottomrule
  \end{tabular}
  \par\vspace{3pt}{\footnotesize\raggedright 499\,K cells, 16~ranks per simulation, 200~iterations, mean $\pm$ population standard deviation over 3~runs; the makespan of a run is the slower of its two simulations, averaged over runs; all times are the solver's ClockTime. Thpt.\ = throughput relative to sequential execution of the same policy's single simulation (C-1P 1249\,s for C-2P, C-1E 1179\,s for C-2E); $T_A/T_B$ = ratio of the two simulations' wall-clock times.\par}
\end{table*}

C-1P lowers one simulation's reservation from 16.0 to 5.9\,vCPU and takes 1249\,s against C-1E's 1179\,s (Table~\ref{tab:results_mux}). This is a 63\% reduction in requested CPU at a 6\% single-job time cost. The requests do not cap execution; the measurements do not isolate whether the difference arises from weights or placement.

Two proportional-request simulations complete in 1412\,s, giving $1.77\times$ throughput relative to sequential C-1P runs. Equal requests give $1.83\times$ relative to their own single-job baseline, with a 1289\,s makespan. The concurrent runs add 9--13\% to single-simulation time, while reserving 11.8\,vCPU under proportional requests or 32\,vCPU under equal requests. The two simulations' completion times are within 10\% of each other: $T_A/T_B$ is 0.94 for C-2P and 1.01 for C-2E. Later-deployed simulations tend to finish slightly later; the slowest-to-fastest ratio stays within 1.08 in the busy-wait sweep. Supplementary Figure~S5 visualizes the two-simulation comparison.

\begin{figure*}[t]
  \centering
  \includegraphics[width=\textwidth]{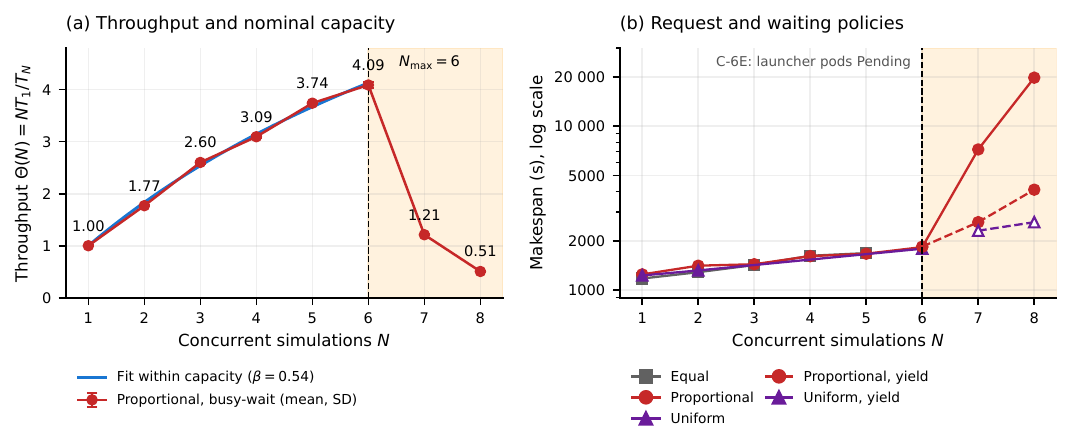}
  \caption{Concurrency on Testbed~B.  (a)~Throughput of the proportional busy-wait sweep (mean $\pm$ standard deviation over runs) with the single-parameter model ($\beta = 0.54$, fitted for $\rho_N \le 1$; it tracks observation within 4\% and is not extended beyond that domain).  Past the nominal capacity (shaded, 112 and 128 worker pods on 96 vCPUs), busy-wait spin displaces computation and throughput collapses to $0.51\times$ at $N=8$. (b)~Makespans of every policy on a logarithmic scale.  Equal, proportional and uniform requests were measured at the densities marked (uniform at $N = 1, 2$ and 6 only; lines guide the eye) and differ by at most 10\% up to $N = 5$; equal requests cannot be scheduled at $N = 6$.  Beyond the boundary yielding waits (dashed) soften the collapse and uniform requests do better than proportional ones.  Hollow markers are single runs; the other points are means over 2--6 runs (Section~\ref{sec:setup_metrics}).}
  \label{fig:pareto-cliff}
\end{figure*}

\begin{table}[t]
  \centering
  \small
  \caption{Equal-request baselines for the concurrency sweep on Testbed~B}
  \label{tab:equal_sweep}
  \begin{tabular}{@{}lrrrr@{}}
    \toprule
    $N$ & \textbf{C-$N$E (s)} & $\Theta_E$ & \textbf{C-$N$P (s)} & $\Theta_P$ \\
    \midrule
    2 & $1289 \pm 11$ & 1.83 & $1412 \pm 9$ & 1.77 \\
    3 & $1425 \pm 9$  & 2.48 & $1441 \pm 16$ & 2.60 \\
    4 & $1623 \pm 9$  & 2.91 & $1615 \pm 15$ & 3.09 \\
    5 & $1684 \pm 8$  & 3.50 & $1670 \pm 6$ & 3.74 \\
    6 & unschedulable & -- & $1833 \pm 22$ & 4.09 \\
    \bottomrule
  \end{tabular}
  \par\vspace{3pt}{\footnotesize\raggedright Makespan (mean $\pm$ population standard deviation; C-$N$E three runs, C-$N$P three runs and six at $N = 3, 4$) and throughput relative to each strategy's own single-simulation baseline (C-1E 1179\,s, C-1P 1249\,s).\par}
\end{table}

Extending co-location from $N=2$ to $N=6$ raises proportional-request throughput from $1.77\times$ to $4.09\times$, while scheduling efficiency, throughput divided by $N$, falls from 88\% to 68\% (Figure~\ref{fig:pareto-cliff}(a)). The largest early drop is between $N=3$ (87\%) and $N=4$ (77\%). For latency-sensitive use, $N=3$ offers a smaller 15\% increase in per-simulation time than the 47\% at $N=6$; the latter gives the highest measured throughput.

Equal requests produce makespans within 2\% of proportional requests at $N=3$--$5$ (Table~\ref{tab:equal_sweep}). Their throughput ratios differ partly because each uses its own single-job reference: against C-1E's 1179\,s, the proportional configurations reach $1.67\times$ at $N=2$ and $3.86\times$ at $N=6$. The request shape does not account for the concurrency gain. Lower totals instead reduce the reserved footprint: at $N=5$, the same jobs request 29.5\,vCPU with proportional reservations against 80\,vCPU with equal reservations, a distinction that also matters under a namespace \texttt{ResourceQuota}.

At $N=6$, the 96 worker pods of C-6E request all 96 allocatable vCPUs, leaving its six 50\,m launcher pods Pending. This is a launcher-budget failure: an equal vector of 990\,m per rank would leave room in request arithmetic, but was not tested. C-6P leaves about 61\,vCPU of reservation headroom and obtains 9\% more throughput than C-5P. Its benefit at this boundary is the reduced reservation, rather than admission that no equal vector could provide. Requests alone would also admit $N=7$ and $N=8$; the worker-count guard prevents that over-admission.

\begin{table}[t]
  \centering
  \small
  \caption{Uniform low requests against proportional requests on Testbed~B}
  \label{tab:uniform_mux}
  \begin{tabular}{@{}lrrrr@{}}
    \toprule
    $N$ & \textbf{C-$N$P (s)} & \textbf{C-$N$U (s)} & \textbf{U vs P} & $\Theta_U$ \\
    \midrule
    1 & $1241 \pm 13$ & $1231 \pm 18$ & $-0.8\%$ & 0.96 \\
    2 & $1388 \pm 18$ & $1321 \pm 10$ & $-4.8\%$ & 1.79 \\
    6 & 1818          & $1792 \pm 11$ & $-1.4\%$ & 3.95 \\
    7 (yield) & 2564 & 2308 & $-10.0\%$ & 3.58 \\
    8 (yield) & 4051 & 2600 & $-35.8\%$ & 3.63 \\
    \bottomrule
  \end{tabular}
  \par\vspace{3pt}{\footnotesize\raggedright C-$N$U: 369\,m on every rank, 5.9\,vCPU per simulation, the job total of the proportional vector; each pair run in one session (makespan; C-$N$P two runs at $N = 1, 2$ and one at $N = 6$--$8$, C-$N$U three runs at $N = 1, 2$, two at $N = 6$ and one at $N = 7, 8$; the runs at $N = 7, 8$ use yielding MPI waits). $\Theta_U$ is relative to the equal-request single simulation C-1E (1179\,s); on the basis of the sweep (C-1P, 1249\,s) the values at $N = 7$ and 8 are 3.79 and 3.84.\par}
\end{table}

The same-budget uniform controls separate total reservation from vector shape (Table~\ref{tab:uniform_mux}). Each proportional/uniform pair was measured in one session, with the same 5.9\,vCPU reservation per job. Uniform requests are within 1\% of proportional requests at $N=1$, 5\% faster at $N=2$ and 1\% faster at $N=6$, where throughput is $3.95\times$ against the equal-request baseline. The proportional anchors reproduce the earlier sweep within 2\%. Thus both shapes admit six simulations and have similar performance within nominal capacity; no consistent advantage of proportional requests was measured. Their behavior beyond that capacity is considered next.

\paragraph{Throughput within nominal capacity.}
Fitting Eq.~\eqref{eq:throughput-model} over $N=2,\ldots,6$ gives $\beta=0.54$ and reproduces the measured points within 4\%, including the boundary prediction of $4.13\times$ against the observed $4.09\times$. Calibration from the $N=2$ point alone gives $\beta=0.78$ and conservative predictions, 8\% low at $N=3$ and at most 12\% low at $N=5$. The fit describes this measured range; it is not extended beyond $\rho_N=1$.

\subsubsection{Concurrency Limits and MPI Waiting}
\label{sec:res_capacity}

\begin{table}[t]
  \centering
  \small
  \caption{Busy-wait versus yielding MPI waits at and beyond the capacity boundary}
  \label{tab:yield}
  \begin{tabular}{@{}lrrrr@{}}
    \toprule
    $N$ & \textbf{Busy-wait (s)} & $\Theta$ & \textbf{Yield (s)} & $\Theta$ \\
    \midrule
    6 & 1833   & 4.09 & $1844 \pm 10$ & 4.06 \\
    7 & 7221   & 1.21 & $2602 \pm 38$ & 3.36 \\
    8 & 19\,783 & 0.51 & $4105 \pm 54$ & 2.43 \\
    \bottomrule
  \end{tabular}
  \par\vspace{3pt}{\footnotesize\raggedright Testbed~B, proportional requests, makespan; busy-wait values are means from the sweep above, run-to-run variation below 2\%; yield runs: two at each $N$, those at $N=7$ and $N=8$ from two sessions.\par}
\end{table}

Once worker ranks outnumber logical CPUs, the throughput trend reverses (Figure~\ref{fig:pareto-cliff}). At $N=7$, 112 workers compete for 96 vCPUs and makespan rises from 1833\,s to 7221\,s, reducing throughput to $1.21\times$. At $N=8$ it falls to $0.51\times$, below the sequential baseline. Mean iteration time rises from 9\,s to 36\,s to 99\,s across $N=6,7,8$. The collapse is repeatable, with run-to-run variation below 2\% in the over-capacity regime and a slowest-to-fastest ratio of at most 1.06.

Node CPU occupancy stays near 95\% on both sides of the boundary and no bandwidth throttling occurs, since these configurations set no limits. The change is consistent with busy-wait spin competing directly with computation when runnable workers exceed logical CPUs (Section~\ref{sec:theory}). On this cluster, the nominal boundary of Eq.~\eqref{eq:capacity-boundary} coincides with the measured turning point.

Yielding waits reduce the cost of crossing that boundary. With \texttt{mpi\_yield\_when\_idle} enabled, the waiting rank gives another runnable thread an opportunity to execute. At $N=6$, yielding changes makespan by less than 1\%, from 1833\,s to 1844\,s (Table~\ref{tab:yield}). At $N=7$, makespan falls from 7221\,s to 2602\,s and throughput rises to about $3.4\times$; at $N=8$, the corresponding values are 19783\,s to 4105\,s and $0.51\times$ to $2.4\times$. Under proportional requests, the extra simulations still do not recover the boundary throughput, and the fairness ratio reaches 1.13 at $N=7$.

Uniform requests perform better in the same-session yielding pairs (Table~\ref{tab:uniform_mux}): 2308\,s at $N=7$ and 2600\,s at $N=8$, 10\% and 36\% below proportional requests, with a fairness ratio of 1.02. Throughput holds at about $3.6\times$ against the equal-request baseline, or $3.8\times$ against the proportional sweep's baseline. These are single-run comparisons at each density. The differences are consistent with low proportional weights delaying a sparse rank's short, critical-path compute phase under contention, although per-core scheduling traces were not collected to confirm this mechanism.

Yielding therefore softens over-admission, but it does not improve on the best throughput measured within the boundary. The uniform yielding results also remain below C-6U on the same throughput baseline. The controller uses the worker-count boundary as an admission guard while adjusting reservations and admitting successive simulations.

\FloatBarrier
\subsection{In-Place CPU Request Adjustment}
\label{sec:res_dynamic}

\begin{figure*}[t]
  \centering
  \includegraphics[width=\textwidth]{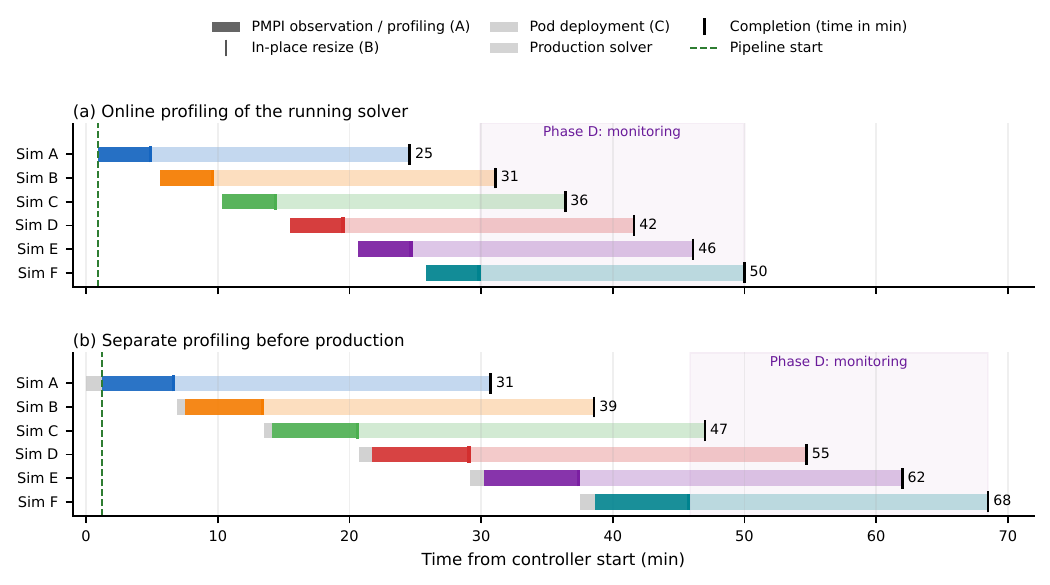}
  \caption{Six-simulation controller runs on Testbed~B, with identical axis ranges measured from each controller's start. (a)~Online observation of the production solver and in-place resizing 241\,s after its start. (b)~A separate 50-iteration profiling pass precedes each production run. Dark spans show Phase~A observation or profiling; vertical resize markers show Phase~B updates; lighter bars show the 200-iteration production runs. The dashed line marks the pipeline origin, the first production start in (a) and first profiling start in (b); initial deployment is excluded from the pipeline durations in Table~\ref{tab:controller_runs}.}
  \label{fig:dynamic-timeline}
\end{figure*}

\begin{table}[!b]
  \centering
  \footnotesize
  \caption{The two end-to-end controller runs on Testbed~B}
  \label{tab:controller_runs}
  \begin{tabular}{@{}lrr@{}}
    \toprule
 & \textbf{Online} & \textbf{Separate pass} \\
    \midrule
    Admission interval & 5.0\,min & 7.8\,min \\
    Sixth start after the first & 24.9\,min & 39.1\,min \\
    Runs in progress (max / mean) & 5 / 3.1 & 4 / 2.2 \\
    Pipeline & 49.1\,min & 67.3\,min \\
    Batch throughput & 2.54$\times$ & 1.86$\times$ \\
    Reservation after Phase~B (six) & 46.2\,vCPU & 47.6\,vCPU \\
    In-place resizes (Phase~B + D) & 96 + 24 & 96 + 12 \\
    Resizes of running solvers & 120 & 12 \\
    Per-simulation wall clock & 1422--1567\,s & 1360--1577\,s \\
    Pod restarts & 0 & 0 \\
    \bottomrule
  \end{tabular}
  \par\vspace{3pt}{\footnotesize\raggedright Six 16-rank, 499\,K-cell simulations; one run per design. Online profiling uses the first 240\,s of production; the separate pass uses 50 iterations. Batch throughput is $6 \times 1249$\,s divided by the pipeline, from the first production start (online) or first profiling start (separate pass) to the last completion, excluding initial deployment. Static C-6P reaches $4.09\times$ with simultaneous starts.\par}
\end{table}

Two controller runs on Testbed~B each admit six 16-rank simulations while estimating and adjusting their reservations (Table~\ref{tab:controller_runs}). The online run profiles the production solver itself; the separate-pass run profiles before production. Across both runs, initial estimates give 140--330\,m for sparse, 360--680\,m for medium and 690--1000\,m for dense ranks, totaling 6.5--8.4\,vCPU per simulation. These budgets exceed the static cell-count vector because measured duty cycles vary less across ranks than cell counts.

\paragraph{Profiling the running solver.}
Every simulation in the online run begins production at equal requests with the PMPI wrapper attached. After 240\,s, the controller uses the latest 60\,s window to resize the running solver's pods (Figure~\ref{fig:dynamic-timeline}(a)). The updates occur 241\,s after solver start, at iterations 33--46 of 200, and reduce each reservation to 6.5--8.3\,vCPU, or 46.2\,vCPU for six simulations. Subsequent windows show continued solver progress, and all simulations finish with their iteration counts unchanged. The controller performs 120 resizes of running solvers, 96 in Phase~B and 24 in Phase~D, with zero pod restarts.

Online observation overlaps with production, avoiding the separate profiling passes of 5.4--7.3 minutes. Admission intervals are 5.0 minutes rather than 7.8, and the six-simulation pipeline takes 49 minutes rather than 67. Batch throughput is $2.5\times$ for the online run against $1.9\times$ for the separate-pass run. Serial admission accounts for part of the remaining gap to static C-6P's $4.09\times$, in which all six simulations start together with a known request vector.

\paragraph{Cost of a separate profiling pass.}
In the separate-pass run, each production solver starts after a 50-iteration profile and a Phase~B resize (Figure~\ref{fig:dynamic-timeline}(b)). Its 108 updates comprise 96 between profiling and production and 12 on running solvers in Phase~D; there are zero pod restarts. The final Phase~B reservation is 47.6\,vCPU. The sixth production run starts 39 minutes after the first, and at most four are in progress at once. Phase~D makes three bounded adjustments, raising the slowest simulation's dense-rank requests by 10\% when the fastest-to-slowest rate ratio exceeds 1.15. This run demonstrates unattended profiling, resizing and admission, with the profiling delay visible in the pipeline.

Each design was evaluated once. These runs establish in-place adjustment of running MPI solvers and the cost of the demonstrated admission process; they do not compare dynamic and static policies under matched arrival sequences and budgets or test adaptation to a physical change in load. Failure handling remains untested (Section~\ref{sec:discussion_scope}).

The auxiliary preset-phase C5 checks also complete without solver disruption. The pitzDaily update fires at iteration 100, whereas the NACA update occurs at approximately 98\% completion and does not establish a phase-adaptive benefit. Supplementary Section~S2 gives the protocol and timings.

\FloatBarrier
\subsection{Rank Allocation at a Fixed Core Budget}
\label{sec:res_blocks}

\begin{table*}[t]
  \centering
  \small
  \caption{Ranks per zone at a fixed budget of 16 physical cores (Testbed~C)}
  \label{tab:blocks}
  \begin{tabular}{@{}lrrrrrrr@{}}
    \toprule
    \textbf{Ranks per zone} & \shortstack{\textbf{Near-wall cells}\\\textbf{per rank (4.0\,M)}} & \multicolumn{2}{c}{\textbf{1.0\,M (s)}} & \multicolumn{2}{c}{\textbf{4.0\,M (s)}} & \shortstack{\textbf{vs 8/4/4}\\\textbf{4.0\,M, 50 it.}} & \shortstack{\textbf{Pressure it./step}\\\textbf{4.0\,M, first 50 it.}} \\
     & & 200 it. & 4000 it. & 50 it. & 400 it. & & \\
    \midrule
    8/4/4            & 680\,166 & 661 & 6048 & 643 & 4306 & baseline & 100 \\
    4/4/8            & 340\,083 & --  & --   & 292 & --   & $-55\%$ & 61 \\
    2/4/10           & 272\,066 & 162 & 2973 & 275 & 1014 & $-57\%$ & 62 \\
    2/2/12           & 226\,722 & --  & --   & 371 & --   & $-42\%$ & 60 \\
    Scotch, equal cells & 249\,395 & 350 & 6459 & 471 & 2554 & $-27\%$ & 125 \\
    \bottomrule
  \end{tabular}
  \par\vspace{3pt}{\footnotesize\raggedright One rank per core. Wall clock for a fixed 200 iterations on the 1.0\,M mesh (3 runs, spread at most 1\,s), for 4000 iterations on the 1.0\,M mesh (one run each of 8/4/4, 2/4/10 and Scotch with the case's residual criterion active; none met it, and the layouts not listed were not run this long), and on the 4.0\,M mesh for 50 iterations (2 runs, spread at most 1\,s) and for 400 iterations (one run each of 8/4/4, 2/4/10 and Scotch). ``Near-wall cells per rank'' is the cell count of one near-wall rank on the 4.0\,M mesh, ``vs 8/4/4'' compares the 4.0\,M times at 50 iterations, and $p$ iterations are pressure-solver iterations per step over the first 50 iterations of the 4.0\,M runs (GAMG with cross-processor agglomeration).\par}
\end{table*}

Partitioning determines the work assigned to the ranks whose CPU reservations are studied above. The complementary Testbed~C comparison changes rank allocation across zones while retaining their cell fractions and a budget of 16 physical cores (Section~\ref{sec:meth_blocks}). Every rank is pinned to its own core in Table~\ref{tab:blocks}. Moving ranks from the far field toward the near-wall zone, from 8/4/4 to 2/4/10, shortens the 4.0\,M-cell run from 643\,s to 275\,s over 50 iterations, a factor of 2.3. The 4/4/8 layout takes 292\,s, while equal Scotch takes 471\,s. This comparison changes the workload and partition geometry rather than the CPU requests of a fixed set of ranks.

\begin{figure}[!t]
  \centering
  \includegraphics[width=\columnwidth]{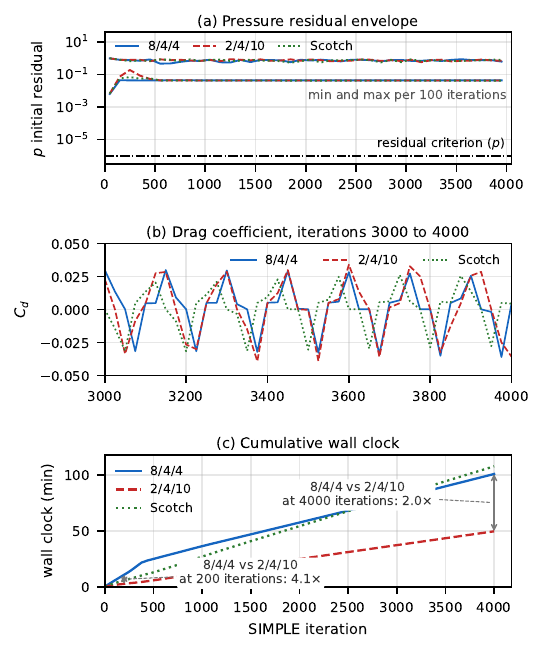}
  \caption{The three 1.0\,M-cell layouts run for 4000 iterations on Testbed~C (one run each).  (a)~Initial pressure residual, shown as its minimum and maximum envelopes over every 100 iterations; the dashed line is the residual criterion of the case, which no layout reaches.  (b)~Drag coefficient over iterations 3000--4000. (c)~Cumulative wall-clock time, with the 200- and 4000-iteration horizons marked: the margin of 2/4/10 over 8/4/4 is 4.1$\times$ at 200 iterations and 2.0$\times$ at 4000.}
  \label{fig:phase15}
\end{figure}

\paragraph{Iteration horizon and solver work.}
The advantage depends on the iteration horizon. On the 1.0\,M mesh, 2/4/10 is 4.1$\times$ faster than 8/4/4 over 200 iterations but 2.0$\times$ over 4000; its settled-phase time per step is 1.7$\times$ lower (Figure~\ref{fig:phase15}). The corresponding margin over Scotch is about 2.2$\times$ at each horizon. On the 4.0\,M mesh, the margin over 8/4/4 grows from 2.3$\times$ at 50 iterations to 4.2$\times$ at 400, while the margin over Scotch grows from 1.7$\times$ to 2.5$\times$.

The two meshes differ because their pressure-solver workloads evolve differently. For 1.0\,M cells, 8/4/4 drops from 86 pressure iterations per step during the first 200 steps to 27 over steps 501--4000; 2/4/10 averages 23 in both intervals. For 4.0\,M cells, the 2/4/10 count falls from 62 to 4.4, whereas 8/4/4 remains above 65 over steps 201--400. None of the layouts reaches the prescribed convergence criterion within the 4000-step 1.0\,M runs or the 400-step 4.0\,M runs. The residual and drag histories continue to oscillate, so these are fixed-iteration performance comparisons. Supplementary Section~S4 provides the detailed interval statistics.

Both the work per rank and the pressure-iteration count change with the partition. On the 4.0\,M mesh, the wall-clock time per pressure iteration falls from 0.129\,s for 8/4/4 to 0.089\,s for 2/4/10, while the pressure iterations per step fall from 100 to 62. These factors describe the measured ratio without isolating its causes: the time per pressure iteration also includes the other solver work amortized over those iterations (Section~\ref{sec:setup_pressure_cost}). Scotch has the lowest such time, 0.075\,s, but needs 125 pressure iterations per step, compared with 60--62 for the refined zone layouts. Allocating still more ranks to the near-wall zone shifts the bottleneck to the medium zone: 2/2/12 takes 371\,s. The 2/4/10 layout is the best of the four zone allocations tested, not an optimum.

\paragraph{Far-field core sharing.}
\begin{table*}[t]
  \centering
  \small
  \caption{Sharing cores among the far-field ranks (Testbed~C)}
  \label{tab:sharing}
  \begin{tabular}{@{}lllrrrr@{}}
    \toprule
    \textbf{Far-field ranks} & \textbf{Waits} & \textbf{Layout} & \textbf{1.0\,M (s)} & \textbf{4.0\,M (s)} & \multicolumn{2}{c}{\textbf{s per $p$ iteration}} \\
     & & & & & 1.0\,M & 4.0\,M \\
    \midrule
    1 per core & busy-wait & 8/4/4  & 661  & 643 & 0.039 & 0.129 \\
    1 per core & yield     & 8/4/4  & 657  & --  & 0.038 & -- \\
    2 per core & yield     & 8/4/4  & 758  & --  & 0.044 & -- \\
    4 per core & yield     & 8/4/4  & 1162 & --  & 0.068 & -- \\
    4 per core & yield     & 8/4/10 & 1280 & 490 & 0.070 & 0.087 \\
    unpinned (18 cores) & yield & 8/4/10 & $960 \pm 158$ & $474 \pm 10$ & 0.053 & 0.084 \\
    1 per core (two ranks) & busy-wait & 2/4/10 & 162 & 275 & 0.036 & 0.089 \\
    \bottomrule
  \end{tabular}
  \par\vspace{3pt}{\footnotesize\raggedright Wall clock for 200 iterations on the 1.0\,M mesh (3 runs) and 50 on the 4.0\,M mesh (2 runs), and the wall-clock time per pressure iteration (run time divided by the pressure iterations performed, Section~\ref{sec:setup_pressure_cost}; it includes the momentum, turbulence and energy work of each step). Spread at most 3\,s except where stated ($\pm$ is the population standard deviation over the runs); the unpinned configuration could use all 18 cores of the host.\par}
\end{table*}

Keeping the 1.0\,M-cell 8/4/4 partition fixed isolates the cost of sharing far-field cores (Table~\ref{tab:sharing}). Yielding alone has little cost, 657\,s against 661\,s. Two far-field ranks per core increase time by 15\% and four by 76\%. Under default busy-waiting, the packed 8/4/10 layout needed 350 to 370\,s per iteration, more than a hundred times the 3.3\,s of the one-rank-per-core 8/4/4 run, and was stopped after three iterations. With identical solver work, four ranks per core add 29\,ms per pressure iteration, from 0.039\,s to 0.068\,s. The increase is consistent with scheduling and synchronization overhead, although per-collective latency was not measured.

For 4.0\,M cells, the shared-core 8/4/10 layout takes 0.087\,s per pressure iteration, close to 2/4/10's 0.089\,s. These layouts differ in partition and iteration count; an unshared 8/4/10 would require 22 cores and was not run. The unpinned exploratory configuration could use all 18 host cores, so it is not a matched 16-core comparison. The shared 8/4/10 layout still needs 112 pressure iterations per step against 62 for 2/4/10: it finishes in 490\,s, 24\% faster than 8/4/4 but 78\% slower than 2/4/10.

For these cases, changing the rank allocation outperforms sharing the original far-field ranks. If the block topology is fixed, yielding allows those ranks to share cores and leaves cores for the near-wall zone, with a relative overhead that decreases as work between synchronizations increases. The CPU multiplexing experiments in Section~\ref{sec:res_duty} retain the original 8/4/4 partition on Testbed~B. They do not use these refined layouts, whose ranks wait less; the two sets of gains have not been measured together.

\subsection{Implications and Limitations}
\label{sec:discussion}

\subsubsection{Implications for Shared-Cluster Scheduling}
\label{sec:discussion_resources}
\label{sec:discussion_scheduling}

CPU multiplexing is governed by both reservation accounting and execution contention. Lowering requests reduces a job's reserved footprint while allowing its ranks to use available CPU time. The single-job and same-budget controls show why total reservation and request shape should be assessed separately. A low PMPI duty cycle motivates a reduced reservation, whose cost can then be checked against single-job time and the fairness of co-located runs. CPU occupancy alone cannot supply this signal when MPI waits busy-poll.

Within the nominal boundary, additional simulations can occupy logical CPUs left unoccupied by a single job. The measurements do not separate the remaining effects of SMT, memory-system sharing and overlapping collectives. At higher density, the worker-count guard becomes necessary even when ample reservation headroom remains. The throughput model calibrated from a dual-simulation run gives a conservative description within that boundary, while the controller itself uses reservation and worker-count checks for admission. On the fixed cluster's on-demand price of \$4.12 per cluster-hour, the measured throughput corresponds to 43--76\% lower amortized instance cost per simulation from $N=2$ to $N=6$ (Supplementary Table~S4). This is a conversion of cluster running time, excluding other cloud costs.

These observations complement earlier co-location work. Iancu et al.\ \cite{iancu2010oversubscription} report both useful sharing and degradation in synchronization-heavy workloads; here the transition is measured alongside Kubernetes reservation accounting. Heracles \cite{lo2015heracles} controls interference through hardware partitioning, whereas this study uses declarative requests without hardware isolation. Gang scheduling \cite{feitelson1992gang}, Volcano and Kueue provide job admission and queueing; the request mechanism supplies the reservations against which that admission operates. With the tested 1000\,m requests, up to five simulations fit while the sixth lacks launcher headroom. The sequential reference is not a measured performance baseline for those schedulers.

Request adjustment also leaves the running solver's partition, rank count and state intact. This makes it complementary to repartitioning and malleability, which change the work assigned to ranks. The Testbed~C comparison evaluates that partition-side choice independently; a joint evaluation is future work. The companion preprint \cite{xie2026multiplexing} reports additional mesh-size, NAS-kernel, node-count, and memory observations beyond the experiments included here.

\subsubsection{Scope and Limitations}
\label{sec:discussion_scope}

\emph{Workloads and scale.} The single-simulation comparisons cover 16\,K and 101\,K cells at 4 and 16 ranks, plus 1.0\,M and 4.0\,M cells at 16--22 ranks on one node. Multiplexing uses the same 499\,K-cell solver and mesh in every simulation, with up to 128 worker pods on 96 vCPUs. All cases are two-dimensional steady RANS; heterogeneous workloads, 3-D hybrid meshes and transient solvers may have different cost distributions. Production CFD can involve $O(10^7)$--$O(10^9)$ cells and hundreds to thousands of ranks, where collective communication can become more costly \cite{hoefler2010characterizing}. Behavior at $\ge$64 ranks per job remains open. Scaling comparisons are restricted to the same testbed.

\emph{Numerical convergence.} Testbed~C's long runs do not reach the prescribed residual criterion, and their performance ratios depend on the iteration horizon (Section~\ref{sec:res_blocks}). The pressure-iteration cost describes the timing difference without isolating a causal contribution from partition geometry, load balance or multigrid convergence.

\emph{Hardware and session variability.} Testbeds~A and~B use SMT instances, so a logical CPU is not an exclusive physical core. The physical hosts and unrecorded SMT-sibling placement differ between sessions. Identical c5.xlarge configurations differ by up to 21\% in wall-clock time across the March, 4 September and 6 September 2026 sessions. The changes are not uniform: C2-16w shifts by 7\% and C4-16L by 21\% between the September sessions, reversing their order (Tables~\ref{tab:naca_16rank} and~\ref{tab:naca_16rank_uniform}). Comparisons are therefore made within a session, and the approximately 10\% request-vector difference in Table~\ref{tab:naca_16rank} is an observation from that session. Memory-bandwidth contention is inferred rather than measured; hardware counters are not exposed on the Nitro-virtualized instances.

\emph{Isolation and recovery.} Co-located simulations share nodes, cores and caches under stock Kubernetes namespaces, cgroups and \texttt{ResourceQuota}. Side-channel risks and interference between mutually untrusted tenants are outside the evaluation. The controller provides no MPI fault recovery; resize rejection, partial deployment, and recovery after node loss remain untested. The two runs lost to worker-node failure (Section~\ref{sec:setup_metrics}) would also have been lost under equal requests. GPU workloads were not evaluated.

\section{Conclusions and Future Work}
\label{sec:conclusions}

\begin{enumerate}
  \item \textbf{CPU reservations can be reduced with a small single-job cost.}  Reducing a job's reservation from 16.0 to 5.9\,vCPU increased single-job wall-clock time by at most 6\%, with either load-proportional or uniform low requests. Per-rank requests set the scheduler reservation and cgroup CPU weight; without hard limits, ranks can use available CPU capacity during computation. In matched controls, decomposition changes reduced wall-clock time by 17--23\%, and removing CPU limits saved 6\%. In a single-run pitzDaily comparison on burstable instances, limits equal to sub-core requests increased wall-clock time about 27$\times$ relative to the same requests without limits.

  \item \textbf{CPU multiplexing increases throughput within a measured concurrency boundary.}  Six co-located simulations on twelve nodes reached $4.09\times$ the throughput of sequential execution with the same reduced requests. A 5.9\,vCPU reservation per job admitted the sixth simulation where the tested 1000\,m per-rank requests left no room for the launchers. Matched controls showed no consistent advantage of proportional request shaping. Admission also had to account for worker counts: at eight simulations, busy-wait throughput fell to $0.51\times$ despite remaining reservation headroom. Yielding waits softened this decline, and uniform requests performed better than proportional requests in the matched single-run comparisons beyond the boundary.

  \item \textbf{Online profiling enables in-place adjustment of running MPI solvers.}  In one six-simulation run, the online controller profiled the production solvers and applied 120 in-place resizes with zero pod restarts. The pipeline completed in 49 minutes and reached $2.5\times$ the throughput of sequential execution. A design using a separate profiling pass, also evaluated once, required 67 minutes. Serial admission accounts for part of the gap to the static $4.09\times$.
\end{enumerate}

For the evaluated workloads, CPU multiplexing required attention to both reserved CPU and runnable worker counts. In-place request updates allowed reservations to change while the MPI solvers and their in-memory state were preserved.

Partitioning determines the workload that these CPU reservations must accommodate. A complementary comparison examined the effect of redistributing ranks across mesh zones. On a single node with a fixed budget of 16 physical cores, reallocating ranks from the far field to the near-wall zone (8/4/4 to 2/4/10) shortened the fixed-iteration runs. Relative to the original concentric allocation, wall-clock time fell by factors of 2.0 to 4.2 across the 1.0\,M- and 4.0\,M-cell meshes, depending on the iteration horizon (Table~\ref{tab:blocks}). The layout also outperformed balanced Scotch partitions. These changes combined a more even share of work across ranks with fewer pressure iterations per step; their contributions were not isolated.

\paragraph{Future work.}
Future work will evaluate CPU multiplexing in three-dimensional reacting flows with detailed chemical kinetics, including co-located simulations with different and time-varying loads. These tests will measure how the balance between local computation and MPI waiting affects throughput, interference, and the useful concurrency range. Repeated comparisons with static policies under matched arrival sequences and resource budgets will assess the value of online reservation adjustment. The CPU configurations will include limits at or above requests and equal requests that leave capacity for launchers.

A joint evaluation of rank allocation and reduced CPU reservations on the same cluster will examine how a better-balanced partition changes the scope for CPU multiplexing. A mesh-based estimator will propose ranks per zone before execution and guide this comparison.

A further direction is coordinated CPU and GPU sharing for combustion CFD, with transport on CPU ranks and neural chemistry inference on shared GPUs. Joint profiling would test whether the timing of these phases permits useful sharing across simulations, and how device memory, GPU queueing and MPI waiting constrain throughput and fairness. Surrogate accuracy and acceleration should be evaluated separately from any gains due to resource sharing.

\section*{Declarations}

\paragraph{Abbreviations.}
CFD, computational fluid dynamics; CFS, Completely Fair Scheduler; CNI, container network interface; EEVDF, Earliest Eligible Virtual Deadline First; EFS, Elastic File System; GAMG, geometric-algebraic multigrid; K8s, Kubernetes; KEP, Kubernetes Enhancement Proposal; MPI, Message Passing Interface; PMPI, MPI profiling interface; QoS, Quality of Service; RANS, Reynolds-averaged Navier--Stokes; SMT, simultaneous multithreading; vCPU, virtual CPU.

\paragraph{Funding.}
This research received no external funding; the cloud experiments were self-funded.

\paragraph{Conflict of interest.}
The author declares no competing interests.

\paragraph{Data availability.}
The experiment scripts, controller, PMPI profiling library, analysis code, Kubernetes manifests, and result tables are maintained in \href{https://github.com/Xieldor/K8s-Paper}{\texttt{Xieldor/K8s-Paper}} (Testbed~A) and \href{https://github.com/Xieldor/K8s-CFD-Multiplexing}{\texttt{Xieldor/K8s-CFD-Multiplexing}} (Testbeds~B and~C); these repositories will be made publicly available upon journal publication, together with the raw solver logs, PMPI traces, and per-pod metric time series. Until then, data and code are available from the author on request.

\paragraph{Author contribution.}
Tianfang Xie is the sole author and performed all conceptualization, implementation, experimentation, analysis, and writing.

\newcommand{\BIBdecl}{\setlength{\itemsep}{1pt}}
\bibliographystyle{IEEEtran}
\bibliography{references}

\clearpage
\onecolumn
\setcounter{section}{0}\setcounter{figure}{0}\setcounter{table}{0}
\renewcommand{\thesection}{S\arabic{section}}
\renewcommand{\thefigure}{S\arabic{figure}}
\renewcommand{\thetable}{S\arabic{table}}
\renewcommand{\theHsection}{supp.\arabic{section}}
\renewcommand{\theHfigure}{supp.\arabic{figure}}
\renewcommand{\theHtable}{supp.\arabic{table}}
\begin{center}{\LARGE\bfseries Supplementary Material}\end{center}

\raggedbottom

\noindent Tables~S1 to~S4 provide the numerical setup of the pitzDaily case (S1; Section~4.2.1 of the main text), the 4-rank configurations (S2; Section~4.3), the 4-rank wall-clock times (S3; Section~5.1.1 and Figure~7(a) of the main text) and the compute-instance cost conversion of the concurrency sweep (S4; Section~5.5.1).  Figure~S1 shows the mesh-generation topology, and Figure~S2 illustrates the fixed-core rank layouts. Sections~S1--S4 and Figures~S3--S5 contain auxiliary execution checks, preset-phase adjustment, two-simulation plots and extended fixed-core statistics. Additional run statistics accompany Section~4.4.3. Configuration names, testbeds and statistical conventions are those of the main text.

\begin{table}[htbp]
  \centering
  \small
  \caption{pitzDaily case setup}
  \label{tab:pitzdaily_bc}
  \begin{tabular}{@{}lp{4.6cm}@{}}
    \toprule
    \textbf{Parameter} & \textbf{Value} \\
    \midrule
    Mesh cells         & 12\,225 (structured \texttt{blockMesh}) \\
    Domain ($x$, $y$)  & $[-0.021,\,0.29]\times[-0.025,\,0.025]$\,m \\
    Step height        & $h = 0.0254$\,m, expansion ratio $= 2.0$ \\
    Solver             & \texttt{simpleFoam} (steady SIMPLE) \\
    Turbulence         & $k$-$\varepsilon$, standard wall functions \\
    $\nu$              & $1\times10^{-5}$\,m$^2$/s \\
    $U_\infty$         & 10\,m/s; $Re_h = 25\,400$ \\
    Inlet $k$          & $0.375$\,m$^2$/s$^2$ ($I=5\%$) \\
    Inlet $\varepsilon$ & $14.855$\,m$^2$/s$^3$ \\
    Pressure BCs       & Outlet $p=0$; inlet zero-gradient \\
    Walls              & No-slip; $k$/$\varepsilon$ wall functions \\
    Relaxation         & $p$: 0.3;\; $U$,\,$k$,\,$\varepsilon$: 0.7 \\
    Convergence        & Residuals $< 10^{-5}$ \\
    \bottomrule
  \end{tabular}
\end{table}

\begin{table}[htbp]
  \centering
  \small
  \caption{4-rank configurations}
  \label{tab:configs}
  \begin{tabular}{@{}lp{1.85cm}p{3.2cm}@{}}
    \toprule
    \textbf{ID} & \textbf{Name} & \textbf{Description} \\
    \midrule
    C0 & hostNetwork    & K8s pods, EC2 native network (no overlay) \\
    C1 & Single-pod     & 4 MPI procs in one pod, shared memory \\
    C2 & K8s equal      & 4 pods, 1.0 vCPU each, equal decomp. \\
    C2-manual & Concentric equal & Concentric decomp., 1.0 vCPU each (NACA) \\
    C3 & Prop. + limits & Weights $(1,1,5,15)$, hard CPU limits \\
    C4 & Prop. req-only & Weights $(1,1,5,15)$, requests only \\
    C4b& Exact prop.    & Exact fractions 182/182/909/2727\,m, req.\ only \\
    C5 & Pseudo-dynamic & Prop., In-Place resize at a preset iteration \\
    \bottomrule
  \end{tabular}
  \par\vspace{3pt}{\footnotesize\raggedright pitzDaily on the t3 cluster, and 16\,K NACA on Testbed~A; the one NACA run of C5, a mechanism demonstration, used the t3 cluster. C3 is the only proportional configuration with CPU limits; C4, C4b and C5 are requests-only.\par}
\end{table}

\begin{table}[htbp]
  \centering
  \small
  \caption{NACA~0012 wall-clock times at 4 ranks (16\,K cells, Testbed~A)}
  \label{tab:naca_wallclock}
  \setlength{\tabcolsep}{4pt}
  \begin{tabular}{@{}llrc@{}}
    \toprule
    \textbf{Config} & \textbf{Decomposition} & \textbf{Wall clock (s)} & \textbf{vs C2} \\
    \midrule
    C2 equal          & Scotch equal & $209.3 \pm 4.0$ & baseline \\
    C2-manual         & Concentric   & $173.0 \pm 0.8$ & $-17\%$ \\
    \textbf{C4 manual}& \textbf{Concentric} & $\mathbf{167.0 \pm 1.4}$ & $\mathbf{-20\%}$ \\
    \bottomrule
  \end{tabular}
  \par\vspace{3pt}{\footnotesize\raggedright 3 runs each; mean $\pm$ population standard deviation. ``Concentric'' is the distance-based manual decomposition.\par}
\end{table}

\begin{table}[htbp]
  \centering
  \caption{Estimated compute-instance cost per simulation on the fixed twelve-node cluster}
  \label{tab:cost}
  \small
  \begin{tabular}{@{}rrrrc@{}}
    \toprule
    $N$ & Total time (s) & Total cost (\$) & \$/sim & Saving \\
    \midrule
    1 & 1249 & 1.43 & 1.43 & baseline \\
    2 & 1412 & 1.62 & 0.81 & 43\% \\
    3 & 1441 & 1.65 & 0.55 & 62\% \\
    4 & 1615 & 1.85 & 0.46 & 68\% \\
    5 & 1670 & 1.91 & 0.38 & 73\% \\
    6 & 1833 & 2.10 & 0.35 & 76\% \\
    \bottomrule
  \end{tabular}
  \par\vspace{3pt}{\footnotesize\raggedright Testbed~B, twelve c5.2xlarge workers plus the t3.medium master; \$4.12 per cluster-hour is the sum of the us-east-2 on-demand prices listed by AWS in September 2026, counted for the makespan only, without storage, network, deployment or tear-down time. The saving is the amortization of cluster time over more simulations, a conversion of the measured throughput rather than a separate measurement.\par}
\end{table}

\clearpage
\begin{figure}[H]
  \centering
  \includegraphics[width=0.60\textwidth]{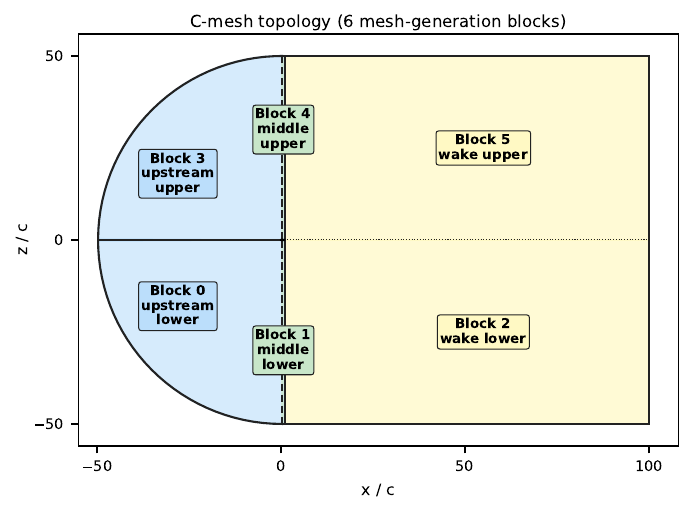}
  \caption{Six-block C-mesh topology used to generate the 498\,834-cell NACA~0012 mesh (zCells = 333, xUCells = 333, xMCells = 138 and xDCells = 278). The block labels identify mesh-generation blocks. MPI rank groups are shown separately in Figure~6 of the main text.}
  \label{fig:mesh-blocks}
\end{figure}

\begin{figure}[H]
  \centering
  \includegraphics[width=0.80\textwidth]{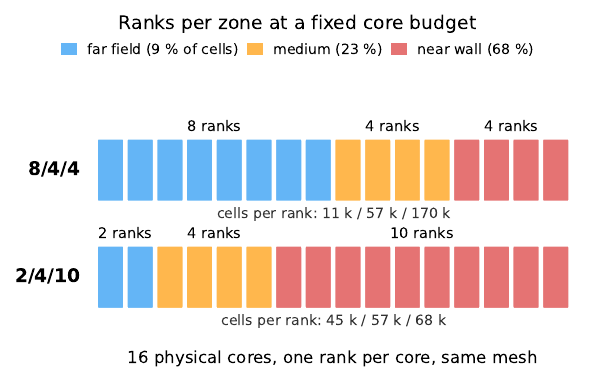}
  \caption{Zone-based rank allocation at a fixed budget of 16 physical cores on Testbed~C (Section~4.3.4 of the main text). The zones retain their cell fractions while the number of ranks per zone changes from 8/4/4 to 2/4/10. Cell counts per rank are rounded values for the 1.0\,M-cell mesh. Each rank is assigned one physical core.}
  \label{fig:architecture}
\end{figure}

\section*{Additional run statistics}
Run-to-run coefficients of variation are below 2\% on Testbed~A and 2.4\% on Testbed~B; the three over-capacity C-7P runs vary by under 2\%.  On Testbed~C the spread is at most 3\,s for pinned configurations; the unpinned configurations vary by $\pm 158$\,s on the 1.0\,M mesh and $\pm 10$\,s on the 4.0\,M mesh.

The Testbed~B campaign contains 274 completed simulation executions: 256 static (121 proportional busy-wait, 45 equal, 27 yielding and 63 in the uniform-comparison session) and 18 under the controller (12 production runs and 6 profiling passes).

The pitzDaily single-run observations span 291--293 iterations and 27--101\,s across configurations, with 2738\,s under hard limits; these timings are used qualitatively.

Numerical consistency is checked within each partition. On Testbed~B, the 200-iteration residual histories are byte-identical across C-1P, C-1U, C-2P, C-2U, C-6P and C-6U. On Testbed~A, configurations sharing a decomposition reach the same convergence criteria in the same number of iterations: 1390 for the equal 101\,K partition and 1449 for weighted Scotch at 16 ranks (Table~8 of the main text).

\FloatBarrier
\section{Auxiliary Single-Simulation Checks}
\label{sec:res_overhead}
The following checks accompany Section~5.1 of the main text. The pitzDaily timings are single-run observations on burstable instances and are used qualitatively.

\begin{figure}[H]
  \centering
  \includegraphics[width=0.75\textwidth]{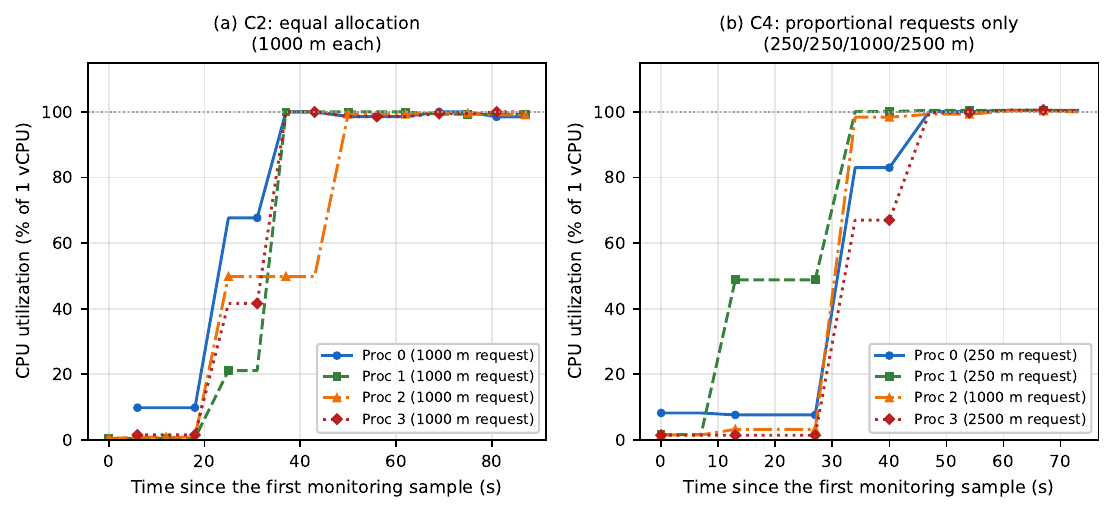}
  \caption{Per-rank CPU utilization over monitoring time (pitzDaily, \texttt{kubectl top} at 5\,s intervals; the origin is the first sample, single runs on burstable instances).  Left: C2 (equal allocation, 1000\,m each).  Right: C4 (proportional requests-only).  After the start-up ramp every rank, including the 250\,m ranks of C4, occupies about one logical CPU: waiting ranks busy-poll, so occupancy does not reveal their idleness.}
  \label{fig:cpu_util}
\end{figure}

Figure~\ref{fig:cpu_util} plots the per-rank CPU utilization sampled via \texttt{kubectl top} at 5\,s intervals for C2 (equal, 1000\,m each) and C4 (proportional requests-only, 250/250/1000/2500\,m); the time origin is the first monitoring sample, and the metrics pipeline lags the solver by a few samples, so the start-up ramp is a monitoring artifact.  After the ramp all four ranks of both configurations sit near 100\% of one logical CPU, the 250\,m ranks of C4 included: what the occupancy signal sees on the small subdomains is busy-wait spin at the barriers, not computation, which is why occupancy cannot tell a waiting rank from a computing one (Section~5.2). Rank~3 of C4 uses 40\% of its 2500\,m request, the remainder of which a single thread cannot consume.

Every 4-rank configuration reserves 4.0\,vCPU, so the reserved CPU time (request multiplied by wall clock) follows the wall-clock times above: C3 holds its reservation 27$\times$ longer than C4, over 10\,000 vCPU-seconds against the 140 of C2.  C4b, the same weights with the exact fractions 182/182/909/2727\,m and no limits, completes in 86\,s against C4's 101\,s (2.5$\times$ and 2.9$\times$ the time of C2): on this 12\,K-cell case the weighted decomposition, whose heaviest rank holds 68\% of the cells, sets the iteration time, and the two request vectors of the same total differ by 15\% in single runs on burstable instances.

A two-node check on the four-rank pitzDaily configuration compares the communication paths of the 4-rank experiments; it is specific to this configuration and is not a network benchmark.  Configuration~C2 runs four pods each with 1.0~CPU request and limit under equal Scotch decomposition, matching C1 in every resource dimension.

\begin{figure}[H]
  \centering
  \includegraphics[width=0.72\textwidth]{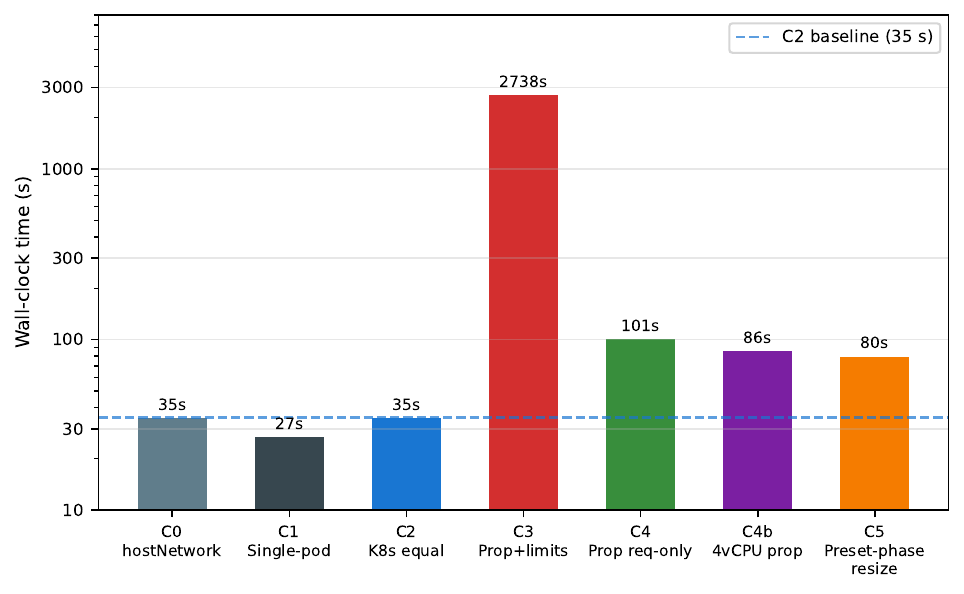}
  \caption{Wall-clock time for the seven 4-rank configurations on pitzDaily (291--293 iterations to convergence; single runs on burstable t3.xlarge instances).  The dashed blue line marks the C2 cross-node K8s baseline (35\,s).  Note the logarithmic scale: C3 (proportional with hard limits) requires 2738\,s, 27$\times$ the time of C4, which has the same decomposition and requests without limits; the 78$\times$ gap to C2 also includes the change of decomposition.}
  \label{fig:wallclock}
\end{figure}

Figure~\ref{fig:wallclock} shows that C2 wall-clock time exceeds C1 by 30\% on pitzDaily (35\,s vs.\ 27\,s; single-run, burstable instances). This gap arises from MPI communication traversing the network rather than shared memory: all four processes in C1 run on one node and communicate via zero-copy shared memory (\texttt{btl=self,vader}), whereas C2 distributes two processes per worker node.  To separate the Kubernetes network layer from cross-node TCP/IP latency, C0 runs the same four processes in two pods (two ranks per pod) with \texttt{hostNetwork:\ true}, bypassing the Flannel VXLAN overlay and using the native EC2 VPC network directly.  C0 also takes 35\,s in this single run.  Because C0 changes two things at once, bypassing the overlay and giving each pair of ranks a shared-memory path inside its pod, the comparison does not isolate the overlay's cost; the 8\,s gap between C1 and the two cross-node configurations is consistent with cross-node TCP/IP communication replacing shared memory. 

The 16-rank NACA control also records a smaller limit effect: the master rank's CPU time is 7\% below its wall-clock time with the 1000\,m limit, against 2\% without it. The proportional vectors correspond to cgroup weights of $8/36/107$ for C4-16 and $3/13/39$ for C4-16L. These observations do not isolate the scheduling mechanism behind the single-session request-vector difference.

\FloatBarrier
\section{Preset-Phase In-Place Adjustment}
\label{sec:supp_preset}
Configuration~C5 applies the protocol of Section~3.4 to a running solver.  The scaler monitors simulation progress via \texttt{kubectl exec} polling of \texttt{processor0/} time directories and patches pod CPU requests at pre-defined phase boundaries using \texttt{kubectl patch -{}-subresource resize -{}-type=json}. On pitzDaily (80\,s total), the Phase~1$\to$2 transition fired at iteration~100 (34.4\% progress), patching all four pods within $\Delta t_{\mathrm{patch}} < 1\,\mathrm{s}$; each \texttt{kubectl patch} completed in $<250\,\mathrm{ms}$ and the kubelet propagated the cgroup update within $\sim$5\,s, about twenty pitzDaily iterations, during which the solver continued. Solver residuals continued decreasing monotonically through the transition with no stall or divergence artifact, consistent with in-place CPU resizing leaving the MPI process, its in-memory field data and inter-pod communication channels intact in this run.

On the longer NACA~0012 case (16\,K cells, four ranks, t3.xlarge cluster; 475\,s for 5000 iterations, concentric decomposition), the transition fired at $\sim$98\% completion, so the phase-adaptive benefit was not measured on this case; C5 completed without solver disruption, which is what the pseudo-dynamic experiment is used to establish.

\FloatBarrier
\section{Two-Simulation Co-location}
\label{sec:supp_two_job}
\begin{figure}[H]
  \centering
  \includegraphics[width=0.52\textwidth]{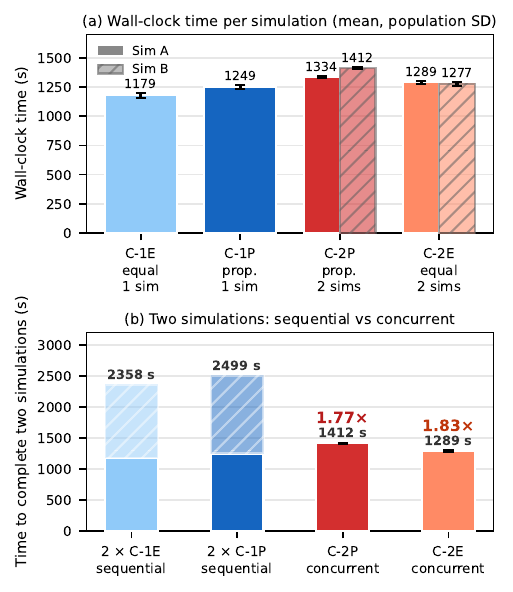}
  \caption{Two-simulation co-location on Testbed~B (values of Table~10 in the main text).  (a)~Wall-clock time per configuration (solid: simulation~A, hatched: simulation~B; mean $\pm$ population standard deviation over three runs).  (b)~Time to complete two simulations, sequential (twice the single-simulation time) against concurrent, with the throughput gain annotated: C-2P $1.77\times$, C-2E $1.83\times$.}
  \label{fig:throughput}
\end{figure}

Figure~\ref{fig:throughput} gives the visual comparison corresponding to Table~10 and Section~5.2.2 of the main text. All times and baselines follow that table.

\FloatBarrier
\section{Extended Fixed-Core Run Statistics}
\label{sec:supp_fixed_core}
\small
The following interval statistics accompany Table~15 and Section~5.4 of the main text.

These are times for a fixed number of outer iterations, and the ratio depends on how many.  One run each of 8/4/4, 2/4/10 and Scotch on the 1.0\,M mesh was carried to 4000 iterations with the case's residual criterion active (Table~15, fourth column; Figure~12 of the main text).  The first 200 iterations are a start-up phase: 8/4/4 needs 86 pressure iterations per step there and 3.2\,s per step; over iterations 501--4000 it averages 27 pressure iterations and 1.30\,s per step, 2/4/10 averages 23 and 0.75\,s, and Scotch 67 and 1.63\,s.  The 4000-iteration run took 2973\,s with 2/4/10, compared with 6048\,s for 8/4/4 and 6459\,s for Scotch.  On the 4.0\,M mesh one run each of the same three layouts was carried to 400 iterations (Table~15, sixth column); their first 50 iterations reproduce the 50-iteration runs (identical pressure-iteration and residual sequences, times within 2\%).  Over iterations 201--400, 8/4/4 averages 65 pressure iterations and 9.17\,s per step, 2/4/10 4.4 and 1.83\,s, and Scotch 43 and 4.29\,s; the 400-iteration run took 1014\,s with 2/4/10, compared with 4306\,s for 8/4/4 and 2554\,s for Scotch.

None of the three layouts reached the prescribed convergence criterion within 4000 iterations.  Their iteration histories differ (after 200 iterations the initial pressure residual is 0.05 under 8/4/4, 0.17 under 2/4/10 and 0.12 under Scotch), but from iteration 500 on all three show sustained oscillations of similar magnitude: the initial pressure residual of each layout cycles between 0.043 and 0.88 with a mean of 0.10, the mean initial residuals of momentum and energy lie within 14\% of one another across the layouts, and the drag coefficient oscillates about zero with a standard deviation of 0.017--0.022 over the last 1000 iterations.  The 4.0\,M runs do not converge within 400 iterations either: over the last 100 iterations the initial pressure residual ranges over 0.03--0.51 under 8/4/4, 0.40--0.74 under 2/4/10 and 0.27--0.76 under Scotch, and the drag coefficient oscillates with a standard deviation of 0.015--0.039.

On the 4.0\,M mesh, the dense ranks of 8/4/4 compute for 94--97\% of the run and the sparse ranks for 10\%; with 2/4/10, the range is 64--94\%. Equal Scotch keeps every rank 90--94\% busy. For 2/2/12, the two medium ranks contain 453\,445 cells each and become the bottleneck, while the dense-rank duty cycle falls to 66\%.

The unpinned 8/4/10 exploration uses 250\,m requests for shared ranks and allows all 18 host cores (allowed core IDs 0--17, with cores 0--1 busy). Its 4.0\,M run takes 474\,s and 0.084\,s per pressure iteration. This observation is separate from the pinned 16-core comparisons in Table~16 of the main text.

\end{document}